\documentclass[a4paper,11pt]{article}
\pdfoutput=1 

\usepackage{jheppub} 

\usepackage[T1]{fontenc} 
\newcommand{\nn}{\nonumber}
\newcommand{\tr}{\mathrm{Tr}\:}

\renewcommand{\Im}{\mathrm{Im}\:}

\renewcommand{\Re}{\mathrm{Re}\:}

\title{\Large \bf Non Abelian T-duality in Gauged Linear Sigma Models}
\author[a,b]{Nana Cabo Bizet}
\author[a,c]{Aldo Mart\'{\i}nez-Merino}
\author[d,e]{Leopoldo A. Pando Zayas}
\author[a,f]{Roberto Santos-Silva}

\affiliation[a]{Departamento de F\'{\i}sica,
Divisi\'on de Ciencias e Ingenier\'{\i}as\\ Universidad de Guanajuato,
 Loma del Bosque 103, 37150,  Le\'on, Guanajuato, M\'exico}
\affiliation[b]{Mandelstam Institute for Theoretical Physics, School of Physics, and National Institute for Theoretical Physics, University of the Witwatersrand, Johannesburg, WITS 2050, South Africa}
\affiliation[c]{Investigador C\'atedra CONACyT, Facultad de Ciencias en F\'{\i}sica y Matem\'aticas, Universidad Aut\'onoma de Chiapas, Carretera Emiliano Zapata Km. 8.0, Rancho San Francisco Col. Ter\'an,
Ciudad Universitaria, C.P. 29050, Tuxtla Guti\'errez, Chiapas, M\'exico.}
\affiliation[d]{ Leinweber Center for Theoretical Physics, 
Randall Laboratory of Physics,\\
 The University of Michigan, Ann Arbor, MI 48109-1120, USA}
\affiliation[e]{The Abdus Salam International Centre for Theoretical Physics\\
Strada Costiera 11, 34014 Trieste, Italy}
\affiliation[f]{Departamento de Ciencias Naturales y Exactas, CUValles, Universidad de Guadalajara. Carretera Guadalajara-Ameca Km. 45.5, C.P. 46600, Ameca, Jalisco, M\'exico}

\emailAdd{nana@fisica.ugto.mx}
\emailAdd{a.merino@fisica.ugto.mx}
\emailAdd{lpandoz@umich.edu}
\emailAdd{roberto.santos@academicos.udg.mx}


\abstract{
Abelian T-duality in Gauged Linear Sigma Models (GLSM) forms the basis of the physical understanding of Mirror Symmetry as presented by Hori and Vafa. We consider an alternative formulation of Abelian T-duality on GLSM's as a gauging of a global $U(1)$ symmetry with the  addition of appropriate Lagrange multipliers. For GLSMs with Abelian gauge groups and without superpotential we reproduce the dual models introduced by Hori and Vafa. We extend the construction to formulate non-Abelian T-duality on GLSMs with global non-Abelian symmetries. The equations of motion that lead to the dual model are obtained for a general group, they depend in general on semi-chiral superfields; for  cases such 
as $SU(2)$ they depend on twisted chiral superfields. We solve the equations of motion for an $SU(2)$ gauged group with a choice of a particular Lie algebra direction of the vector superfield. This direction covers a non-Abelian sector that can be described by a family of Abelian dualities. The dual model Lagrangian depends on twisted chiral superfields and  a twisted superpotential is generated. We explore some non-perturbative aspects by making an Ansatz for the instanton corrections in the dual theories. We verify that  the effective potential for the $U(1)$ field strength in a fixed configuration on the original theory matches the one of the dual theory. Imposing restrictions on the vector superfield, more general non-Abelian dual models are obtained.  We analyze the dual models via the geometry of their susy vacua. }

\begin{document}
\maketitle
 \flushbottom

\tableofcontents
\section{Introduction}

String theory is arguably the field of physics that has stimulated the most work in  various fields of mathematics. One of the most prolific example is mirror symmetry. From the physical point view, mirror symmetry is an example of the power of symmetries and dualities in string theory. Indeed, the deepest physical explanations for mirror symmetry are rooted in the idea of duality, more specifically, T-duality. One of the first suggestions that mirror symmetry can be understood in the language of T-duality was put forward in \cite{Morrison:1995yh}; the insightful work of Strominger, Yau and Zaslow further developed this idea in a particular context \cite{Strominger:1996it}. The state-of-the-art description of mirror symmetry presented in  \cite{Hori:2000kt,Hori:2003ic} ties together many works around the concept of T-duality, including non-perturbative effects. 

A practical way to understand T-duality is to consider a system with a global $U(1)$ symmetry, promotes this symmetry to a local symmetry by introducing a gauge field and a Lagrange multiplier enforcing flatness of the corresponding connection. Then, the T-dual models are obtained as the result of integration in different orders  \cite{Giveon:1994fu}.  A similar proposal was put forward for the case of non-Abelian T-duality  \cite{delaOssa:1992vci}. There has recently been a revival in implementations of non-Abelian T-duality (NATD)  in the context of supergravity solutions fueled by an understanding of its implementation in the RR sector  \cite{Sfetsos:2010uq,Lozano:2011kb}. NATD has been widely applied as a solution generating technique for some supergravity backgrounds and has provided many new examples of pairs in the context of the AdS/CFT correspondence (see \cite{Itsios:2017cew,vanGorsel:2017goj} and references therein).   Given the mounting evidence in favor of NATD as a symmetry of  supergravity with RR fields, and indirectly of a class of non-linear sigma models in string theory, it is natural to explore the possibility that NATD could be implemented at the level of 2D gauged linear sigma models (GLSM). Indeed, there is a well-established connection between GLSM and NLSM in the context of string theory.

In this manuscript we study NATD in GLSM. We have various motivations in mind. For example, we hope that  further understanding of NATD in GLSM could lead to a clarification of mirror symmetry in more general Calabi-Yau varieties \cite{Hori:2006dk,Caldararu:2007tc}, specially in determinantal varieties  \cite{Hori:2011pd,Jockers:2012zr}.  Here we work on the implementation of non-Abelian
T-duality, planning the future exploration of the precise connections to mirror symmetry. We follow the standard definition of T-duality stated above and implement it for a class of GLSM with a non-Abelian global symmetry.  The process is carried out  by taking a non-Abelian global symmetry in a gauge theory and gauging it by the introduction of an extra non-Abelian gauge field. Then a Lagrange multiplier term is added (as in the original formulation of the duality) to enforce flatness of the corresponding connection. Eliminating the Lagrange multiplier
or the gauged field leads to the original or the dual model depending on which field is integrated out first. We illustrate this procedure first in the case of a scalar field minimally coupled to a gauge field.  Later we study several examples of Abelian T-duality in GLSMs.  The main idea consists on integrate the vector superfield obtaining a set of equations of motion to express the original coordinates, chiral superfields, in terms of the dual coordinates, which are  twisted chiral superfields. This dualization method has the advantage with respect to the method of Hori and Vafa \cite{Hori:2000kt} that allows a straightforward generalization to non-Abelian symmetries. 

Then we formulate non-Abelian T-duality for a GLSM $U(1)$  theory with a generic global symmetry acting on chiral superfields. We gauge the symmetry by introducing a vector superfield and add a Lagrangian term with an un-constrained superfield as Lagrange multiplier. Integrating the vector superfield one arrives to a set of equations of motion leading to the dual theory. The dual superfields satisfy a semichiral condition, but in special cases such as $SU(2)$ it reduces to a twisted chiral condition. To explore the duality further  we consider as a model a GLSM with two equally charged chiral superfields, with a global $SU(2)$ symmetry. We study the model in detail, dualizing the $SU(2)$ symmetry, to obtain the dual K\"ahler potential and the twisted superpotential.  First we focus on a family of Abelian dualities along an $SU(2)$ vector superfield direction which turns out to be more general because the dual theories describe a non-Abelian set of the $SU(2)$ vector superfield. Then we consider a restricted vector superfield that leads to a fully non-Abelian dual model.

A crucial ingredient that was missing in the original attempts to cast mirror symmetry as T-duality was the inclusion of nonperturbative effects  such as instantons; this was addressed systematically in the work of Hori and Vafa  \cite{Hori:2000kt}. Motivated by this approach we also take some steps in this direction and include corrections to the twisted superpotential and demonstrate that it satisfies some of the expected properties as that it leads to coinciding twisted effective superpotentials for the gauge field strength in the original and the dual model. In addition, by dualizing an Abelian direction in $SU(2)$
and considering the instanton twisted superpotential one obtains the same theory as the one obtained via a $U(1)$
dualization.

The manuscript is organized a follows: In Section \ref{Sec:Review} we start reviewing supersymmetric gauge theories in 2D with (2,2) supersymmetry.  In Section \ref{Sec:AbTD} we present Abelian T-Duality in GLSMs.  In \ref{Subsec:Ex} we show an example of gauging a global symmetry in a gauge theory, which is given by two scalars fields minimally coupled to a $U(1)$ gauge field. In \ref{Subsec:GLSM} we present a (2,2) 2D gauged linear sigma model with two chiral superfields. In subsection \ref{Subsec:1csf} we describe Abelian T-duality in the mentioned GLSM by first dualizing a theory with a single $U(1)$  along the phase of one chiral superfield. Then we consider the case of a single $U(1)$ gauge group and multiple chiral superfields, together with all the
superfields dualizations. We finish the Section by considering Abelian T-duality of a theory with multiple $U(1)$s and multiple chiral superfields. We demonstrate in subsection \ref{Subsec:multiple} that our approach leads to the same dual model as in the case of Hori and Vafa \cite{Hori:2000kt} .  In subsection \ref{Sec:NATDeom} we formulate non-Abelian T-duality for GLSM with a $U(1)$ gauge symmetry and a generic global symmetry group, obtaining the equations of motion that lead to the dual model. We discuss the generalities of the dual model obtained by dualizing an $SU(2)$ symmetry in a $U(1)$ gauge theory with two equally charged chiral superfields. Subsection  \ref{Subsec:T1}  focuses on a particular case of dualization along a  single $SU(2)$ generator; the following subsection \ref{Subsec:Ta} investigates a dualization along a generic Abelian direction inside of the $SU(2)$ group. Section \ref{Sec:SU2NA} considers a more general dualization for a selection of the vector superfield, which is a truly non-Abelian direction. Moreover, we analyze the vacuum manifold of the dual theory without instantons, which has the same geometry as the vacuum for a family of Abelian symmetries. Section  \ref{Sec:Analysis}  compares  the mirror symmetric theory obtained by dualizing via Abelian T-duality in the scheme of Hori-Vafa  \cite{Hori:2000kt}  with our Abelian dualization, which leads to the same results.  In Section  \ref{Sec:Analysis} as well we compare the dual theories of a family of Abelian dualities inside of $SU(2)$ with the one of a single $U(1)$ dualization, finding a perfect match. We present our conclusions and point to a number of interesting follow up directions and open problems in Section \ref{Sec:Conclusions}. Finally, in a series of Appendices \ref{appA} to \ref{appF} we present supporting calculations and adress some technical questions.

\section{$(2,2)$ Supersymmetric gauge theory in 2 dimensions}\label{Sec:Review}
\label{Sec:Review}

In this Section we recall some aspects of superfield representations in 2D which are well understood from the 4D point of view; our goal is to set up the notation and describe the basic fields which are the content of the GLSMs. The treatment of this Section follows closely the monograph ``Superspace or One Thousand and One Lessons in Supersymmetry" \cite{Gates:1983nr} and Witten's work on GLSMs \cite{Witten:1993yc}, to which the reader is referred for more details. Along the Section we will describe chiral, twisted chiral and vector superfields, and also twisted chiral field strengths.

Let us consider the dimensional reduction of a 4D $\mathcal{N}=1$ supersymmetric gauge theory to 2D, this theory is supersymmetric with 2 supersymmetries: $\mathcal{N}=2$.   In one way of realizing the dimensional reduction denoted $(2,2)$ there are two left $Q_-,\bar Q_-$ and two right $Q_+,\bar Q_+$ supersymmetry generators in 2D  \cite{Witten:1993yc}, coming from the 4 SUSY generators
in 4D. In another way one has only two right supersymmetries,
denoted  $(0,2)$. Here only the operators $Q_+,\bar Q_+$ generate the 2D SUSY \cite{Witten:1993yc}, this case will not be considered in the present work.  We employ the $\mathcal{N}=1$ 4D superfield language 
to describe (anti-)chiral, twisted (anti-)chiral and vector superfields. Twisted chiral superfields arise in
the supersymmetric theory in 2D. In the context of non-linear sigma models there are as well semi-chiral superfields, which constitute representations of (2,2) SUSY \cite{Bogaerts:1999jc,Gates:2007ve}. In this manuscript we are interested on the class of GLSMs considered by Hori-Vafa in \cite{Hori:2000kt} where the representations  of the original theory are given by chiral superfields. However in our description
of non-Abelian T-dualities we will encounter semi-chiral conditions for superfields (see Section \ref{Sec:SU2NA}).

Let us write the supersymmetric covariant derivatives in 2D as\footnote{They come from dimensional reduction
of the 4d covariant derivatives $\bar D_{\dot\alpha}=-\frac{\partial}{\partial \bar\theta^{\dot \alpha}}-i \theta^{\alpha}\sigma_{\alpha\dot \alpha}^{\mu}\partial_{\mu}$
and $D_{\alpha}=\frac{\partial}{\partial \theta^{ \alpha}}+i \bar\theta^{\dot \alpha}\sigma_{\alpha\dot \alpha}^{\mu}\partial_{\mu}$ \cite{Witten:1993yc}.}
\begin{eqnarray}
\bar D_{\pm}&=&-\frac{\partial}{\partial \bar\theta^{\pm}}+i \theta^{\pm}(\partial_0\mp\partial_3),   \; \;  \;   \;   \;
D_{\pm}=\frac{\partial}{\partial \theta^{\pm}}-i \bar \theta^{\pm}(\partial_0\mp\partial_3).\label{covD}
\end{eqnarray}
They will also be denoted as $\bar D_{\dot \alpha}$ and $D_{\alpha}$, with indices $\dot\alpha=+,-$ and $\alpha=+,-$. The superspace coordinates are
denoted as $(x^{\mu},\theta^{\alpha},\bar\theta^{\dot\alpha})$, where the 2D space-time is given by $x^0,x^3$,
and the extra coordinates are $x^1,x^2$.
The covariant derivates (\ref{covD}) are relevant in the restrictions on a generic superfield that lead to the susy representations. In the expressions only the 2D coordinates $x^0$ and $x^3$ are relevant in the dimensional reduction from 4D, because the fields are considered as independent of $x^1, x^2$. The chiral ($\Phi$) and anti-chiral  ($\bar\Phi$)  superfields are defined as:
\begin{eqnarray}
\bar D_+\Phi=\bar D_-\Phi=0, \, \,  D_+\bar\Phi=D_-\bar\Phi=0.\label{Chiralsf}
\end{eqnarray}
Let us also define twisted chiral ($\Psi$) and twisted anti-chiral ($\bar \Psi$) superfields as:
\begin{eqnarray}
\bar D_+\Psi=D_-\Psi=0, \, \,  D_+\bar\Psi=\bar D_-\bar\Psi=0.\label{tChiral}
\end{eqnarray}
These are supersymmetric representations that only occur in 2 dimensions.

A non-Abelian transformation for a chiral superfield $\Phi$ in the fundamental representation of a non-Abelian group is given by:
\begin{eqnarray}
\Phi'=e^{i \Lambda}\Phi,\, \, \Lambda=\Lambda^A T_A,\, \, \bar{D}_{\dot\alpha}\Lambda=0,\label{PhiT}
\end{eqnarray}
where $\Lambda$ generates the transformation and $T_A$ are the generators of the group. The transformation parameters $\Lambda^A$ depend in general on superspace coordinates and $\Lambda$ takes values in the Lie algebra of the group spanned by the generators $T_A$'s.
The last condition in (\ref{PhiT}) on $\Lambda$ makes it a chiral superfield and ensures that the transformed field $\Phi'$ is also chiral. 
Correspondingly, an antichiral superfield $\bar{\Phi}$ transforms as 
\begin{equation}
\bar{\Phi}'=\bar{\Phi}e^{-i \bar{\Lambda}},\, \, \bar{\Lambda}=\bar{\Lambda}^A T_A, \, \, D_{\alpha}\bar{\Lambda}=0,\label{PhiBT}
\end{equation}
where $\bar{\Lambda}$ represents an anti-chiral superfield depending on superspace coordinates. For local transformations, where $\Lambda(x)$ depends on $x^{\mu}$,  one has that $\Lambda\neq \bar{\Lambda}$ and a gauge field is introduced to make the action
covariant. More precisely, the theory has a multiplet of real, Lie algebra valued, scalar superfields $V=V^A  T_A$ transforming under
gauge transformations $\Lambda$ and $\bar \Lambda$ as
\begin{eqnarray}
e^{V'}=e^{i\bar\Lambda} e^V e^{-i \Lambda}.\label{VT}
\end{eqnarray}
From the transformations (\ref{PhiT}) (\ref{PhiBT}) and (\ref{VT}) one sees that the term in the action $\int d\theta^4 \bar\Phi e^V \Phi$ is covariant.

There are covariant derivatives with respect to $\Lambda$ transformations (gauge chiral representation) and with respect to $\bar{\Lambda}$ transformations (gauge antichiral representation). A set of derivatives covariant with respect to $\Lambda$ transformations is given by $\bar{\mathcal{D}}_{\dot \alpha}=\bar{D}_{\dot\alpha}$ and $\mathcal{D}_{\alpha}=e^{-V}D_{\alpha}e^V$. A set of covariant derivatives with respect to $\bar\Lambda$ transformations is given by $\bar{\mathcal{D}}'_{\dot \alpha}=e^V\bar{D}_{\dot\alpha}e^{-V}$ and $\mathcal{D}'_{\alpha}=D_{\alpha}$. These two sets of covariant derivatives allow us to define a 2D gauge invariant field strength, as a twisted chiral superfield, as follows: 
\begin{equation}
\Sigma=\frac{1}{2}\{\bar{\mathcal{D}}_+,\mathcal{D}_{-}\},\, \,  \,  \, \mathcal{D}_+\Sigma=\bar{\mathcal{D}}_{-}\Sigma=0,
\end{equation}
where $\mathcal{D}_{\alpha}$ and $\bar{\mathcal{D}}_{\dot\alpha}$ are the gauge covariant derivatives. The gauge transformation of the field strength is given by
$\Sigma\rightarrow e^{i\Lambda}\Sigma e^{-i \Lambda}$. Analogously, one can construct the field strength
$\bar\Sigma=\frac{1}{2}\{\bar{\mathcal{D}}'_-,\mathcal{D}'_{+}\}$ which is a $\bar\Lambda$ covariant
twisted antichiral superfield. One can write
\begin{eqnarray}
\Sigma&=&\frac{1}{2}\bar{D}_{+}(e^{-V} D_{-} e^V),\label{fieldS}\\
&=&\frac{1}{2}(\bar{D}_{+} e^{-V} D_{-} e^V +e^{-V}\bar{D}_{+} D_{-} e^V).\nn
\end{eqnarray}
The hermitian conjugate of the field strength $\Sigma$ is given by 
\begin{eqnarray}
\bar \Sigma=\frac{1}{2}\{\bar{\mathcal{D}}'_-,\mathcal{D}'_{+}\}=\frac{1}{2} D_+( e^V\bar D_{-} e^{-V})=\frac{1}{2} (D_+ e^V\bar D_{-} e^{-V}+e^VD_{+}\bar D_{-} e^{-V}),
\end{eqnarray}
and has gauge transformation $\bar\Sigma\rightarrow e^{i\bar\Lambda}\bar\Sigma e^{-i \bar\Lambda}$. For an Abelian gauge group one gets the following expression
\begin{eqnarray}
\Sigma_0=\frac{1}{2} \bar D_+ D_- V_0,
\end{eqnarray}
and $\bar \Sigma_0=\frac{1}{2} \bar D_- D_+ V_0$. We denote this field strength with subindex $0$ because it
represents the original $U(1)$ gauge group of the GLSM, while $V$ and $\Sigma$ denote the
gauged global symmetry.

We finish this Section by presenting the expansions of twisted (anti-) chiral superfields and twisted field strengths of an Abelian vector field $V_0$ up to total derivatives. They are given by \cite{rocek84}
\begin{eqnarray}
X_i&=&x_i+\sqrt{2}\theta^+ \bar \chi_+ +\sqrt{2}\bar \theta^{-}\chi_-+2 \theta^+\bar \theta ^- G_i+...,\label{twistedF}\\
\bar X_i&=&\bar x_i+\sqrt{2}\chi_+\bar\theta^+  +\sqrt{2}\bar \chi_-\theta^{-}+2   \theta ^-\bar\theta^+  \bar G_i+...,\nn\\
\Sigma_0&=& \sigma_0+ i\sqrt{2}\theta^+ \bar \lambda_+-i \sqrt{2}\bar \theta^-  \lambda_- + 2 \theta^+\bar \theta ^- (D-i v_{03})+...\nn,\\
\bar \Sigma_0&=& \bar \sigma_0- i\sqrt{2} \lambda_+\bar\theta^++i \sqrt{2}\bar\lambda_-\theta^-  + 2 \theta ^- \bar\theta^+(D+ i v_{03})+...\nn
\end{eqnarray}
The ellipsis represents derivatives of fields contributing to the kinetic terms of the Lagrangian. Looking at the
twisted chiral superfields expansions: $x_i$
is an scalar field, $(\bar\chi_+,\chi_-)$ are spin $\frac{1}{2}$ fermions and $G_i$
are auxiliary fields. All the components fields depend on the space-time coordinates $x^0,x^3$. In the field strength expansion one has the scalar $\sigma_0$, the spin $\frac{1}{2}$ fermions $(\bar\lambda_+,\lambda_-)$, the auxiliary field $D$, and the 2D gauge field strength $v_{03}$. In Appendix
\ref{notation} we collect useful formulas for the superfield calculations performed in the work, in particular we give the conventions
for integration in superspace.

\section{Abelian T-Duality in GLSMs}
\label{Sec:AbTD}

In this Section we formulate Abelian T-duality in gauged linear sigma models as a gauging of a global Abelian symmetry. First, in the bosonic context, we  highlight the differences of this approach with the more traditional T-duality in GLSM
explained in detail in \cite{Hori:2003ic}.
We present the approach of gauging a global symmetry in a theory which already has local symmetries. 
We then consider the case of a GLSM with two chiral fields, a $U(1)$ gauge group, and a global
$U(1)$ symmetry. Then we describe the dualization of all Abelian global symmetries
for a GLSM with a $U(1)$ gauge symmetry and $n$ chiral superfields $\Phi_i$
of charge $Q_i$.  Next, we describe the dualization procedure
for a theory with $m$ $U(1)$ gauge fields along the phase of $n$ chiral superfields. This method applies for more general cases some of which we consider in the following Sections, as the dualization of non-Abelian global symmetries
in GLSMs.

\subsection{Dualization of a gauge theory with global symmetries}
\label{Subsec:Ex}

In this Section we describe the T-dualization of  a theory with two scalar fields coupled to a gauge field with minimal coupling, which possesses an $U(1)$ global symmetry. This example
serves to illustrate the procedure of describing T-duality as a gauging 
of a global symmetry in a gauge theory. The global symmetry $U(1)$ acts as a translation on the scalar fields. The dualization of a scalar field minimally coupled to a gauge field is discussed
in \cite{Hori:2000kt} via another procedure, which consists of constructing
a Lagrangian that after integration of certain of its coordinates gives the original or the dual model. 
Our goal in this Section is to show that the dualization of a gauge theory can be realized by gauging
the global symmetry and adding a Lagrange multiplier term; this approach follows the prescription of 
de la Ossa and Quevedo \cite{delaOssa:1992vci}. 

Consider the theory of two scalar fields, $\phi_1$ and $\phi_2$, minimally coupled to a $U(1)$ gauge field
$A_{\mu}$. This theory has the Lagrangian
\begin{eqnarray}
L_{0}=-\rho^2(\partial_{\mu}\phi_1+Q_1 A_\mu)^2-\rho^2(\partial_{\mu}\phi_2+Q_2 A_\mu)^2,\label{minGauge}
\end{eqnarray}
with $Q_{1}$ and $Q_2$ the charges under the $U(1)$ gauge symmetry. The theory has gauge symmetry given by
the transformations $\phi_{1,2}\rightarrow \phi_1+Q_{1,2}\Lambda, A_\mu\rightarrow A_\mu-\partial_\mu \Lambda$,
with local parameter $\Lambda$. There is as well a global symmetry acting as translations on $\phi_1$ and $\phi_2$ which
can be parametrized as $\phi_1\rightarrow \phi_1+\tilde Q_1 \tilde \Lambda, \phi_2\rightarrow \phi_2$.
This symmetry can be gauged by making the parameter $\tilde\Lambda$ dependent on space-time coordinates.
Thus a gauge field $\tilde A_\mu$ needs to be added. Under this scheme one can construct the Lagrangian
\begin{eqnarray}
L_1=-\rho^2(\partial_{\mu}\phi_1+Q_1 A_\mu+\tilde Q_1 \tilde A_\mu)^2-\rho^2(\partial_{\mu}\phi_2+Q_2 A_\mu)^2
+\psi \tilde F_{\mu\nu}\epsilon^{\mu\nu}.\label{Ltot1}
\end{eqnarray}
Integrating the Lagrange multiplier field $\psi$, i.e., imposing its equation of motion (e.o.m.) one obtains the pure gauge restriction $\tilde F_{\mu\nu}\epsilon^{\mu\nu}=0$,
which applied on (\ref{Ltot1}) leads to the original model (\ref{minGauge}).
Instead, taking a variation of (\ref{Ltot1}) with respect to $\tilde A_{\mu}$, that is, considering the
equation of motion for $\tilde A_{\mu}$, we obtain:
\begin{eqnarray}
\partial^\mu\phi_1+Q_1 A^\mu+\tilde Q_1\tilde A^\mu=-\frac{1}{\rho^2 \tilde Q_1}\partial_\nu\psi \epsilon^{\nu\mu},
\end{eqnarray}
which upon substitution in (\ref{Ltot1}) it leads to the dual Lagrangian:
\begin{eqnarray}
L_{dual}&=&\frac{D-1}{\rho^2}\left(\frac{2}{\tilde Q_1}-1\right)(\partial_{\mu}\psi)^2-\frac{Q_1}{\tilde Q_1}\psi \epsilon^{\mu \nu}F_{\mu \nu}\\
&-&\rho^2(\partial_{\mu}\phi_2+Q_2 A_\mu)^2.\nn
\end{eqnarray}
$D$ represents the space-time dimension. By making the redefinition $\psi\rightarrow 2 \psi/\tilde Q_1$ and fixing $\tilde Q_1$ vs. $D$ to ensure
$(D-1)\left(\frac{2}{\tilde Q_1}-1\right)=-\frac{1}{4}$, one gets the  normalization obtained by Hori-Vafa in their duality procedure \cite{Hori:2000kt}:
\begin{eqnarray}
L_{dual}'&=&-\frac{1}{4 \rho^2}(\partial_{\mu}\psi)^2-Q_1\psi \epsilon^{\mu \nu}\frac{1}{2}F_{\mu \nu}
-\rho^2(\partial_{\mu}\phi_2+Q_2 A_\mu)^2.
\end{eqnarray}
This simple example shows that T-dualities can be described in gauge theories, as a gauging
of remnant global symmetries. This procedure is equivalent to taking as a starting point a more general Lagrangian that encodes the model and the dual model \cite{Hori:2000kt}. It is important, however, to notice that in this case one requires the existence of an extra scalar field which we call spectator (because it is not charged under the gauged global symmetry) to allow for an extra global Abelian symmetry that can be gauged. We will encounter similar situations  in more general cases.

\subsection{Supersymmetric GLSM}
\label{Subsec:GLSM}

Here we describe 2D (2,2) supersymmetric GLSM's with a single $U(1)$ and multiple chiral superfields.
We give the Lagrangian containing the kinetic terms and gauge interactions, without a superpotential. Also we describe the twisted superpotential with its instanton corrections. This theory is the starting point for most of the dualities that will be studied in this work.

The Lagrangian of a 2D (2,2) GLSM with gauge group $U(1)$ and $n$ chiral superfields $\Phi_i$ with charges $Q_i$ can be written as \cite{Hori:2000kt}
\begin{eqnarray}
L_0=\int d^4\theta \left(\sum_{i=1}^n \bar\Phi_ie^{2 Q_i V_0}\Phi_i-\frac{1}{2 e^2}\bar{\Sigma}_0\Sigma_0 \right)-\frac{1}{2} \int d^2\tilde{\theta} t\Sigma_0+c.c.,\label{Lglsm}
\end{eqnarray}
where $t=r-i\theta$, with $r$ the Fayet-Iliopoulos (FI) parameter and $\theta$ the theta angle; $e$ denotes the $U(1)$ gauge coupling. The classical theory is invariant under vector(V) times axial-vector(A) R-symmetries
$U(1)_V\times U(1)_A$, where $\Sigma$ has charge
$(0,2)$. The chiral superfields transform under the vector times axial-vector R-symmetries, the $U(1)_V$ transformation is given by
$\Phi(x,\theta^{\pm},\bar\theta^{\pm})\rightarrow e^{i\alpha q^i_V}\Phi(x,e^{-i \alpha}\theta^{\pm},e^{i \alpha}\bar\theta^{\pm})$,
and the $U(1)_A$ is given by $\Phi(x,\theta^{\pm},\bar\theta^{\pm})\rightarrow e^{i\alpha q^i_A}\Phi(x,e^{\mp i \alpha}\theta^{\pm},e^{\pm i \alpha}\bar\theta^{\pm})$,
with $q_V$ and $q_A$ charges with respect to the vector and axial-vector symmetries.  
The $U(1)_A$ symmetry has a chiral anomaly which is canceled for the charge
relation: $\sum_i Q_i=0$. As we are considering simple examples meant as building blocks
for more realistic cases we do not impose that condition.

In addition, the theory (\ref{Lglsm}) has other global symmetries, which are at least $(N-1)$ $U(1)$ symmetries, those are the phase rotations of the $N$ chiral superfields modulo $U(1)$ gauge transformations. Each charged chiral superfield $\Phi_i$ can be dualized
using the phase rotation symmetry. This gives rise to Abelian T-duality.

Considering one-loop effects, one can obtain an effective superpotential for the vector field $\Sigma_0$
with a fix configuration \cite{Witten:1993yc}. This is done by computing corrections to the vacuum expectation value of the $D$
term. The effect can be interpreted as an effective value for the FI term depending on the scalar component of
$\Sigma_0$, which is $\sigma$. The effective twisted
superpotential reads \cite{Witten:1993yc,Hori:2000kt}
\begin{eqnarray}
W_{eff}(\Sigma_0)=\sum_i -\Sigma_0 Q_i\left( \ln\left(\frac{Q_i\Sigma_0}{\mu}\right)-1\right)-t\Sigma_0.\label{effW}
\end{eqnarray}
This effective superpotential will serve for comparison with the one obtained for
$\Sigma_0$ in the T-dual model.

\subsection{Abelian T-duality}
\label{Subsec:1csf}

Here we describe the T-Duality of a 2D supersymmetric (2,2) GLSM with Abelian gauge group $U(1)$ and two chiral superfields
with charges $Q_1$ and $Q_2$.\footnote{They could be chosen to satisfy the anomaly cancellation condition $Q_1+Q_2=0$.} This  system has one global $U(1)$ symmetry realized as the phase rotations of $\Phi_1$ 
and $\Phi_2$ modulo the $U(1)$ gauge transformations. On the target space parametrized by the scalar components
of $\Phi_{1,2}$ it constitutes a system with one cyclical coordinate.
The procedure we perform in this Section is equivalent to the one presented in \cite{Hori:2000kt}, but consists
of gauging the global symmetry in the Lagrangian and adding a Lagrange multiplier in the scheme 
presented in \cite{Giveon:1993ai}.

Let us fix the gauge to remove the phase transformation of $\Phi_2$
and then implement the T-Duality by gauging the phase rotation of the field $\Phi_1$. Namely
under the Abelian global symmetry with parameter $\alpha$ the fields transform as
\begin{eqnarray}
\Phi_1\rightarrow e^{i \alpha} \Phi_1,\,  \,   \,  \,\Phi_2\rightarrow \Phi_2. \label{transfoGlobal}
\end{eqnarray}
We promote the global symmetry to a local one by introducing a vector
superfield $V$ and the Lagrange multipliers  superfields $\Psi$ and $\bar \Psi$
\cite{Rocek:1991ps}. The Lagrangian reads
\begin{eqnarray}
L_1&=&\int d^4\theta \left(\bar\Phi_1 e^{2 Q_{0} V_0+2 Q V}\Phi_1+\bar\Phi_2 e^{2 Q_{0,2} V_0}\Phi_2+\Psi\Sigma+\bar{\Psi}\bar{\Sigma}-\frac{1}{2 e^2}\bar{\Sigma}_0\Sigma_0\right)\label{lab}\\
&-&\frac{1}{2}\int d^2\tilde{\theta} t\Sigma_0+c.c.,\nn
\end{eqnarray}
where $V_0$ and $\Sigma_0=\frac{1}{2} \bar D_+ D_- V_0$ are the vector superfield and field strength of the $U(1)$ gauge symmetry,
while $V$ and $\Sigma=\frac{1}{2} \bar D_+ D_- V$ are the vector superfield and field strength of the gauged symmetry. We have gauged the global $U(1)$ symmetry via the vector superfield $V$ and the use of the unconstrainted superfields $\Psi$ and $\bar\Psi$ which constitute Lagrange multipliers. Integrating the last two one gets the condition $\Sigma=0$, which is a pure gauge field, leading to the original GLSM
of two chiral superfields coupled to a $U(1)$ vector superfield $V_0$.  The equation of motion obtained by varying the action
w.r.t to $V$, $\frac{\delta S}{\delta V}=0$, is given by
\begin{eqnarray}
\bar\Phi_1 e^{2 Q_{0} V_0+2 Q V}\Phi_1=\frac{\Lambda+\bar\Lambda}{2 Q},\label{eomAb}
\end{eqnarray}
with $\Lambda=\frac{1}{2}\bar D_+ D_- \Psi$ and $\bar\Lambda=\frac{1}{2}\bar D_- D_+  \bar\Psi$. The definition determines $\Lambda$ and $\bar\Lambda$ as twisted
chiral and anti-twisted chiral superfields. From (\ref{eomAb}) one can write
\begin{eqnarray}
V=\frac{1}{2 Q} \ln \frac{\Lambda+\bar\Lambda}{2 Q}-\frac{1}{2 Q}\ln \bar\Phi e^{2 Q_0 V_0}\Phi.
\end{eqnarray}
Plugging the expression for $V$ into (\ref{lab}), the Lagrangian can be written as
\begin{eqnarray}
\label{Eq:L1}
L_1&=&\int d^4 \theta\left( -\frac{\Lambda+\bar\Lambda}{2 Q}\ln \left(\frac{\Lambda+\bar\Lambda}{2 Q}\right)
-\frac{\Lambda+\bar\Lambda}{2 Q}\ln (\bar\Phi_1e^{2 Q_0 V_0}\Phi_1)+\bar\Phi_2 e^{2 Q_{0,2} V_0}\Phi_2-\frac{1}{2 e^2}\bar{\Sigma}_0\Sigma_0\right)\nn
\\
&-&\frac{1}{2}\int d^2\tilde{\theta} t\Sigma_0+c.c.
\end{eqnarray}

Let us work out the second term in the  previous equation (\ref{Eq:L1}), it reads
\begin{eqnarray}
-\int d^4 \theta \frac{\Lambda+\bar\Lambda}{2 Q}(2 Q_0 V_0)&=&
-\int d^4 \theta\frac{Q_0}{Q} (\Lambda V_0)-\int d^4 \theta\frac{Q_0}{Q}( \bar\Lambda V_0)\\
&=&\frac{Q_0}{2 Q}\int d\theta^+d\bar{\theta}^{-}\Sigma_0\Lambda+\frac{Q_0}{2 Q}\int d\bar\theta^+d\theta^{-}\bar\Sigma_0\bar\Lambda.\nn
\end{eqnarray}

This gives the dual Lagrangian
\begin{eqnarray}
L_1&=&\int d^4 \theta\left( -\frac{\Lambda+\bar\Lambda}{2 Q}\ln \left(\frac{\Lambda+\bar\Lambda}{2 Q}\right)
-\frac{\Lambda+\bar\Lambda}{2 Q}\ln (\bar\Phi_1\Phi_1)+\bar\Phi_2 e^{2 Q_{0,2} V_0}\Phi_2-\frac{1}{2 e^2}\bar{\Sigma}_0\Sigma_0\right)\nn
\\
&+&\frac{1}{2}\left(\int d^2\tilde{\theta}( \Lambda Q_0/Q-t)\Sigma_0+\int d^2\tilde{\bar\theta}( \bar \Lambda Q_0/Q-\bar t)\bar\Sigma_0\right).\label{abl}
\end{eqnarray}
The second term in (\ref{abl}) has an extra piece with respect to the one obtained in equation
(3.16) of \cite{Hori:2000kt}. However, this extra piece vanishes due to the following property:
\begin{eqnarray}
\int d^4 \theta\frac{\Lambda}{2 Q}\ln (\bar\Phi_1\Phi_1)&=&\frac{1}{4}\int d\theta^+d\bar\theta^{-}
\frac{\Lambda}{2Q}{\bar D}_+D_{-}\ln (\bar\Phi_1\Phi_1)=0,
\end{eqnarray}
which is fulfilled as a result of $D_{+} \bar\Phi_1=0$ and $D_{-} \bar\Phi_1=0$. Similarly the $\bar\Lambda$ term vanishes. The final dual Lagrangian  can be written as
\begin{eqnarray}
L_1&=&\int d^4 \theta\left( -\frac{\Lambda+\bar\Lambda}{2 Q}\ln \left(\frac{\Lambda+\bar\Lambda}{2 Q}\right)+\bar\Phi_2 e^{2 Q_{0,2} V_0}\Phi_2-\frac{1}{2 e^2}\bar{\Sigma}_0\Sigma_0\right)\label{ablOK0}
\\
&+&\frac{1}{2}\left(\int d^2\tilde{\theta}( \Lambda Q_0/Q-t)\Sigma_0+\int d^2\tilde{\bar\theta}( \bar \Lambda Q_0/Q-\bar t)\bar\Sigma_0\right).\nn
\end{eqnarray}
For $Q=1$ it reproduces the dual Lagrangian obtained with the Hori-Vafa procedure, see 3.1 of \cite{Hori:2000kt}:
\begin{eqnarray}
L_1&=&\int d^4 \theta\left( -\frac{1}{2}(\Lambda+\bar\Lambda)\ln \left(\frac{\Lambda+\bar\Lambda}{2}\right)+\bar\Phi_2 e^{2 Q_{0,2} V_0}\Phi_2-\frac{1}{2 e^2}\bar{\Sigma}_0\Sigma_0\right)\label{ablOK}
\\
&+&\frac{1}{2}\left(\int d^2\tilde{\theta}( \Lambda Q_0-t)\Sigma_0+\int d^2\tilde{\bar\theta}( \bar \Lambda Q_0-\bar t)\bar\Sigma_0\right).\nn
\end{eqnarray}
There is, however, a difference: the presence of the spectator chiral superfield $\Phi_2$. From this Lagrangian and using the expansions of the superfields in (\ref{twistedF}), one can compute explicitly the scalar potential, whose components are given below
\begin{eqnarray}
&&\frac{1}{2} \left( \int d^2 \tilde\theta (\Lambda Q_0 - t) \bar{\Sigma}_0 + \int d^2 \tilde{\bar{\theta}}(\bar{\Lambda} Q_0 - \bar{t}) \Sigma_0  \right) = \nn \\  
&=&(Q_0 y-t) (D-i v_{03})+ \sigma_0 G+ (Q_0 \bar{y} - \bar{t})(D+i v_{03}) +\bar{\sigma}_0 \bar{G} \nn \\
&-&\frac{1}{2e^2} \int d^4 \theta \bar{\Sigma}_0 \Sigma_0 = \frac{1}{2e^2}(D-i v_{03})(D+i v_{03})=\frac{1}{2e^2}(D^2+v^2_{03}), \nn \\
&-&\frac{1}{2} \int d^4 \theta (\Lambda + \bar{\Lambda}) \ln \left(\frac{\Lambda+\bar{\Lambda}}{2}\right)= K_{i \bar{j}} G_{i} \bar{G}_{\bar{j}}, \nn \\
&&\int d^4 \theta \, \bar\Phi_2 e^{2 Q_{0,2} V_0}\Phi_2 = F_2 \bar{F_2} + Q_{0,2} D x_2\bar{x}_2, 
\end{eqnarray}
where $F_2$ and $x_2$ are the auxiliary and scalar components of the superfield $\Phi_2$, $y$ and $G$ are the scalar and auxiliary components of the twisted chiral superfield $\Lambda$, and $i,j=1$ are indices denoting the single twisted superfield with $K_{1\bar 1}$ being the K\"ahler metric evaluated in the scalar $y$. In Appendix \ref{twistedExpansion} we compute generically the scalar potential and interactions for a given K\"ahler potential and twisted superpotential of twisted-chiral and twisted-antichiral superfields. The scalar potential can be written as:
\begin{equation}
U= -\left[Q_0(y+\bar{y}) -(t+\bar{t})\right] D - 2 \Re[ \sigma_0 G] - \frac{1}{2e^2} D^2+K_{i \bar{j}} G_{i} \bar{G}_{\bar{j}}-F_2 \bar{F_2} - Q_{0,2}D x_2\bar{x}_2.
\end{equation}
The contribution of the auxiliary field $F_2$ to the scalar potential is zero since is not coupled to any field. Next, integrating out the auxiliary field $D$ we find
\begin{eqnarray}
D &=& -e^2\left[Q_0(y+\bar{y}) -(t+\bar{t})\right]  -e^2Q_{0,2} x_2 \bar{x}_2.
\end{eqnarray} 
Substituting the previous expression into the potential and reducing terms we obtain the expression 
\begin{equation}
U= \frac{e^2}{2}  \left[ Q_0(y+\bar{y}) -(t+\bar{t}) + Q_{0,2} x_2 \bar{x}_2 \right]^2 - 2 \Re[ \sigma_0 G]+K_{i \bar{j}} G_{i} \bar{G}_{\bar{j}}
\end{equation}
Moreover, we want to eliminate the auxiliary field $G$. Using the equation of motion $K_{1 \bar{1}} G= \bar{\sigma_0}$, the scalar potential reduces to
\begin{equation}
\label{AbScPo}
U= \frac{e^2}{2}  \left[ Q_0(y+\bar{y}) -(t+\bar{t}) + Q_{0,2}x_2 \bar{x}_2 \right]^2 - K^{-1}_{1 \bar{1}} | \sigma_0|^2.
\end{equation}
This potential has three susy vacua at $\sigma_0=0$ and $(Q_0(y+\bar{y}) -(t+\bar{t})) + Q_{0,2} x_2 \bar{x}_2=0$, $x_2=0, (y+\bar{y}) =(t+\bar{t})/Q_0$ or 
$x_2=\sqrt{(t+\bar{t})/Q_{0,2}}, (y+\bar{y})=0 $. Going to the Higgs branch with vacuum $\sigma_0=x_2=0$ and $(y+\bar{y}) =(t+\bar{t})/Q_{0,2}$ the gauge field strength $v_{03}$
gets a large mass in the IR limit, thus it can be integrated out \cite{Hori:2000kt}. Let us integrate the field strength $v_{03}$, the effective scalar potential is given by the sum of $U$ and the terms proportional to $v_{03}$, i.e.
\begin{equation}
U_{eff}=\frac{e^2}{2}  \left[Q_0(y+\bar{y}) -(t+\bar{t})\right]^2  + i \left[Q_0(y-\bar{y})-(t-\bar{t}) \right] v_{03} - \frac{1}{2 e^2} v_{03}^2,\label{UeffAb}
\end{equation}
integrating out $v_{03}$ we get the following expression
\begin{equation}
v_{03}= i e^2 \left[Q_0(y-\bar{y})-(t-\bar{t}) \right].\label{v03Ab}
\end{equation}
Substituting this into (\ref{UeffAb}) we obtain
\begin{eqnarray}
U_{eff}&=& \frac{e^2}{2} (Q_0(y+\bar{y}) -(t+\bar{t}))^2  - \frac{e^2}{2}(Q_0(y-\bar{y}) -(t-\bar{t}))^2 , \\
&=&  2 e^2 (t - Q_0 y) (\bar{t} - Q_0 \bar{y})= 2 e^2 |t - Q_0 y|^2. \nn
\end{eqnarray}
The supersymmetric vacuum lies at the locus $Q y=t$. The susy vacuum arises by taking the Higgs branch, i.e., $\sigma_0=0$ and integrating $v_{03}$ in the IR limit.  This procedure is equivalent to integrating the twisted field strength $\Sigma$
in the twisted superpotential (\ref{ablOK}) to obtain the relation $\Lambda Q_0-t=0$, which implies the previous relation for the scalar components $y$. The resulting theory
comes from incorporating the instanton contribution \cite{Hori:2000kt} to the total twisted superpotential, which can be written as $W_{tot}=\frac{1}{2}( \Lambda Q_0-t)\Sigma_0+e^{-\Lambda}$.  If we look at $W_{tot}$ and integrate out $\Lambda$ this will give rise to the same effective potential for $\Sigma_0$ as in the original theory (\ref{effW}).  If we integrate $\Sigma_0$ the condition $\Lambda Q_0-t=0$  leads to a constant twisted-superpotential $\widetilde W=e^{-t/Q_0}$,
having a theory with a zero scalar potential. We will compare this result with the dual models constituting a family of Abelian T-dualities of Section \ref{Sec:NATD}, which constitute a subset of a non-Abelian T-duality.

The K\"ahler metric of the dual theory reads
\begin{eqnarray}
\partial_{\Lambda}\partial_{\bar\Lambda} K=-\frac{1}{2(\Lambda+\bar\Lambda)}.
\end{eqnarray}
Writing the scalar components of the superfields $\Phi$ as
$\phi=\rho e^{i\theta}$ and $\Lambda$ as $\lambda=\tilde{\rho}+i\tilde{\theta}$, the equation of motion
(\ref{eomAb}) implies the relation $\rho^2=\tilde{\rho}/Q$.  One can describe distances in the dual system with the metric
\begin{eqnarray}
d s^2&=&\frac{d\tilde\rho^2+d\tilde\theta^2}{4 \mathcal{\tilde\rho}},\label{ds2}\\
&=& \frac{1}{Q}d\rho^2+\frac{1}{4 Q\rho^2}d\tilde\theta^2.\nn
\end{eqnarray}
The measure (\ref{ds2}) implies the condition $\Re(\lambda)=\tilde\rho>0$, which after renormalization is corrected to
$\Re(\lambda)\geq-\ln(\Lambda_{UV}/\mu)$, with $\Lambda_{UV}$ the UV cutoff and $\mu$ the energy scale \cite{Hori:2000kt}. The dual fields $\Lambda$ and $\bar \Lambda$ in (\ref{ablOK}) are twisted chiral and twisted anti-chiral superfields.
Therefore, the duality exchanges the chiral superfield $\Phi_1$ by the twisted chiral superfield $\Lambda$.

\subsection{Multiple chiral superfields}
\label{multipleSF}
In this subsection we apply the method to describe T-duality by gauging multiple global symmetries in a gauge theory 
to the case of a 2D  (2,2) GLSM with one $U(1)$ gauge symmetry, with vector superfield and field strength $V_0$ and $\Sigma_0$ and  $N+1$ chiral superfields $\Phi_i$ with charges $Q_i$. This theory has
$U(1)^N$ global symmetries that can be gauged, and from the resulting theory the gauge fields can be  integrated out
to lead to a dual model. This case has also been studied in the work by Hori-Vafa via their dualization method,
here we show the equivalence to the T-duality procedure we are spousing. 

We start with the Lagrangian (\ref{Lglsm}) with $n=N+1$ chiral superfields. We consider a parametrization of the $U(1)^N$ global symmetries, such
that each field $\Phi_i,\, i=1,...,N$ is only charged under the $U(1)_i$ with charge $\hat Q_i$, i.e., under
$U(1)_i$ the chiral superfields transform as:
\begin{eqnarray}
U(1)_i: \Phi_i\rightarrow e^{i \hat Q_i \Lambda_i}\Phi_i, \, \, \, \forall_{j\neq i}\Phi_j\rightarrow \Phi_j.
\end{eqnarray}
The field $\Phi_{N+1}$ is uncharged under the $U(1)^N$ global symmetries. The gauging of these symmetries into the Lagrangian
is given by
\begin{eqnarray}
L&=&\int d^4\theta \left(\sum_{i=1}^N \bar\Phi_ie^{2 Q_i V_0+2\hat Q_i V_i}\Phi_i-\frac{1}{2 e^2}\bar{\Sigma}_0\Sigma_0+\sum_{i=1}^N(\Psi_i\Sigma_i+\bar{\Psi}_i\bar{\Sigma}_i) \right)\label{LglsmM}\\
&+&\int d^4\theta \left( \bar\Phi_{N+1}e^{2 Q_{N+1} V_0}\Phi_{N+1}\right)-\frac{1}{2} \int d^2\tilde{\theta} t\Sigma_0+c.c.,\nn
\end{eqnarray}
where $V_i$, $\Sigma_i$ and $\Psi_i$ denote respectively the vector superfield, field strength and Lagrange multiplier superfield associated to the gauged symmetry $U(1)_i$. The field $\Phi_i$ with $i=1,...,N$ has charge $\hat Q_i$ under the gauged symmetry. $\Phi_{N+1}$ is a spectator superfield, which is not charged under $U(1)_i$, is not dualized, and serves to provide the original theory with an extra $U(1)$ global symmetry. 

The equation of motion obtained by making the variation of the action
w.r.t to $V_i$ as $\frac{\delta S}{\delta V_i}=0$  is given by
\begin{eqnarray}
\bar\Phi_i e^{2 Q_{i} V_0+2 \hat Q_i V_i}\Phi_i=\frac{\Lambda_i+\bar\Lambda_i}{2 \hat Q_i},\label{eomAbi}
\end{eqnarray}
from it one can obtain an expression of the vector superfield of the gauged symmetry as
\begin{eqnarray}
V_i=\frac{1}{2 \hat Q_i} \ln \frac{\Lambda_i+\bar\Lambda_i}{2 \hat Q_i}-\frac{1}{2 \hat Q_i}\ln \bar\Phi_i e^{2 Q_i V_0}\Phi_i.\label{ViM}
\end{eqnarray}
Using (\ref{ViM}) in (\ref{LglsmM}) below, i.e., integrating each of the vector superfields $V_i$ one gets the dual Lagrangian
\begin{eqnarray}
L_{dual}&=&\int d^4 \theta\left( -\sum_{i=1}^N\frac{\Lambda_i+\bar\Lambda_i}{2 \hat Q_i}\ln \left(\frac{\Lambda_i+\bar\Lambda_i}{2 \hat Q_i}\right)+\bar\Phi_{N+1} e^{2 Q_{N+1} V_0}\Phi_{N+1}-\frac{1}{2 e^2}\bar{\Sigma}_0\Sigma_0\right) \nn
\\
&+&\frac{1}{2}\int d^2\tilde{\theta}\left(\sum_{i=1}^N \Lambda_i Q_i/\hat Q_i-t\right)\Sigma_0+\frac{1}{2}\int d^2\tilde{\bar\theta}\left(\sum_{i=1}^N \bar \Lambda_i Q_i/\hat Q_i-\bar t\right)\bar\Sigma_0.\label{LdualM}
\end{eqnarray}
This constitutes a replicated version of (\ref{ablOK0}). It gives an alternative procedure to performing the dualization made by Hori-Vafa \cite{Hori:2000kt} (see their subsection 3.2.1), without the need
of deforming the $D$-term i.e. avoiding the procedure they called in that scheme {\it localization}. More importantly, our procedure
also allows to obtain directly the Lagrangian leading to the dual theories, once the global symmetries to gauge are defined. The equation (\ref{LdualM}) fits their dual model by setting $\forall_i \hat Q_i=1$. The total twisted superpotential, after adding the instanton
contributions reads
\begin{eqnarray}
\widetilde W&=&\left(\sum_{i=1}^N \Lambda_i Q_i-t\right)\Sigma_0+\mu\sum_i e^{-\Lambda_i}.\label{WtwistedM}
\end{eqnarray}

The procedure can be applied as well to more general GLSMs, with a larger gauge symmetry and larger global symmetries.
In particular, it can be applied to the case of a GLSM with multiple $U(1)$ gauge fields. We also expect that it can be applied to GLSMs with non-Abelian gauge symmetries as the ones described in \cite{Jockers:2012zr}; these extensions would be the subject of future work.

\subsection{Multiple $U(1)$s}\label{Subsec:multiple}

In this subsection we perform the dualization procedure along the direction of the global Abelian symmetries in the system
of $N$ chiral superfields and $M$ $U(1)$ gauge symmetries. One requires an extra set of $M$ spectators
fields in order to have $U(1)^N$ global symmetries not contained in the $U(1)^M$ gauge symmetry.
The result of dualization is precisely the dual model presented in Section 3.3 of \cite{Hori:2000kt} plus the
neutral fields under the gauged symmetries, which do not play a role in the duality, but act as spectator 
superfields which allow for the existence of extra global Abelian symmetries.

The chiral superfields $\Phi_i$ have charge $\hat Q_i$ under the gauged
symmetry and charges $Q_{i,a}$ under the original $U(1)_a$ gauge symmetries of the GLSM.  The transformation
parameters with respect to the gauge symmetry and the gauged symmetry read $\Lambda_i$ and $\hat \Lambda_{a}$.
Both transformations are given by
\begin{eqnarray}
U(1)_i&:& \Phi_i\rightarrow e^{i \hat Q_i \Lambda_i}\Phi_i, \, \, \, \forall_{j\neq i}\Phi_j\rightarrow \Phi_j,\\
U(1)_a&:& \Phi_i\rightarrow e^{i \hat Q_{a,i} \hat\Lambda_a}\Phi_i.\nn
\end{eqnarray}

We start with a Lagrangian where the global symmetries have been gauged, and Lagrange multiplier fields have been added:
\begin{eqnarray}
L&=&\int d^4\theta \left(\sum_{i=1}^N \bar\Phi_ie^{\sum_a 2 Q_{i,a} V_a+2\hat Q_i V_i}\Phi_i-\sum_a \frac{1}{2 e_a^2}\bar{\Sigma}_a\Sigma_a+\sum_{i=1}^N(\Psi_i\Sigma_i+\bar{\Psi}_i\bar{\Sigma}_i) \right)\label{LmultU1}\\
&+&\int d^4\theta \left(\sum_{k=N+1}^{M+N} \bar\Phi_{k}e^{\sum_a 2 Q_{k,a} V_a}\Phi_{k}\right)-\sum_a\frac{1}{2} \int d^2\tilde{\theta} t_a\Sigma_a+c.c.,\label{LglsmM2}\nn
\end{eqnarray}
where $V_i$ and $\Sigma_i$ denote the vector superfield and field strength associated to the gauged $U(1)_i$ symmetry, $V_a$ and 
$\Sigma_a$ denote the vector superfield and field strength associated to the GLSM gauge symmetry $U(1)_a$. There are $M$
{\it spectators} chiral superfields $\Phi_k$ with charges only under the original gauge group, and uncharged
with respect to the gauged symmetries.
The equation of motion obtained by varying the action
with respect to $V_i$ as $\frac{\delta S}{\delta V_i}=0$  is given by
\begin{eqnarray}
\bar\Phi_i e^{\sum_a 2 Q_{ia} V_a+2 \hat Q_i V_i}\Phi_i=\frac{\Lambda_i+\bar\Lambda_i}{2 \hat Q_i}.\label{eomAbM}
\end{eqnarray}
In the above equation $\Lambda_i=\frac{1}{2}\bar D_+D_-\Psi_i$. As solution for the $U(1)_i$  vector superfield we obtain the expression
\begin{eqnarray}
V_i=\frac{1}{2 \hat Q_i} \ln \frac{\Lambda_i+\bar\Lambda_i}{2 \hat Q_i}-\frac{1}{2 \hat Q_i}\ln \bar\Phi_i e^{\sum_a 2 Q_{ia} V_a}\Phi_i.\label{ViMult}
\end{eqnarray}
Plugging (\ref{ViMult}) into (\ref{LmultU1}) the dual Lagrangian reads
\begin{eqnarray}
L_{dual}&=&\int d^4 \theta\left( -\sum_{i=1}^N\frac{\Lambda_i+\bar\Lambda_i}{2 \hat Q_i}\ln \left(\frac{\Lambda_i+\bar\Lambda_i}{2 \hat Q_i}\right)+\sum_{k=N+1}^{M+N} \bar\Phi_{k}e^{\sum_a 2 Q_{k,a} V_a}\Phi_{k}-\sum_a\frac{1}{2 e_a^2}\bar{\Sigma}_a\Sigma_a\right)\nn
\\
&+&\sum_a\frac{1}{2}\int d^2\tilde{\theta}\left(\sum_{i=1}^N \Lambda_i Q_{ia}/\hat Q_i-t_a\right)\Sigma_a+h.c.\label{LdualMult}
\end{eqnarray}
Let us compare now with the dualization procedure of Hori-Vafa. Setting the charges under the gauged symmetries as $\hat Q_i=1$,
the total twisted superpotential, after adding the instanton contributions reads
\begin{eqnarray}
\widetilde W&=&\sum_a \Sigma_a\left(\sum_{i=1}^N \Lambda_i Q_{ia}-t_a\right)+\mu\sum_i e^{-\Lambda_i}.\label{WtwistedMult}
\end{eqnarray}
This coincides with the results of Section 3.3  in Hori-Vafa work \cite{Hori:2000kt} but we employ the approach of gauging
the remnant global symmetries.  In this case the dual manifolds obtained by this procedure are the same as in the
Hori-Vafa cases, because the twisted superpotential in the effective dual theories coincide. The dual theory is obtained by taking
the Higgs branch of vacua where the
vev of the scalar components of $\Phi_k$ is $\phi_k=0$, the vev of the scalar components of $\Sigma_a$ is $\sigma_a=0$ and the field
strength $v_{03,a}$ is integrated. In this case all the $\Sigma_a$ are integrated out, to give the ralation between the fields $\sum_{i=1}^N \Lambda_i Q_{ia}=t_a$ and the twisted superpotential $\widetilde W=\mu\sum_i e^{-\Lambda_i}$. This implies that the Abelian dualization
reproduces the same results for Mirror Symmetry as the Hori-Vafa dualization. In Section \ref{Sec:Analysis} we analyze this Abelian duality in more detail for the case discussed in the text of two chiral superfields with equal charges and
a single $U(1)$ writing the scalar potential in the dual theory.  We have tested in this Section the method of gauging a global symmetry
in GLSMs, in order to describe Abelian T-dualities. Next, we are going to see the advantage of this method in implementing
non-Abelian dualities in GLSMs.

\section{Non Abelian T-Duality in GLSMs}
\label{Sec:NATD}
In this Section we implement NATD in a 2D (2,2) supersymmetric GLSM. 
We start with a GLSM with Abelian gauge group $U(1)$, $N$ chiral superfields 
$\Phi_{k,i}$  with charges $Q_k$, $\sum_k n_k=N$, $i=1,...,n_k$ where $n_k$ is the number of chiral superfields with charge
$Q_k$. Due to the coincidence of charges there is a group of global non-Abelian symmetry which acts on the chiral superfields.  The non-Abelian global symmetry, denoted by $G$, can be gauged incorporating to the action the
vector superfield $V$ and adding a Lagrange multiplier $\Psi$. When this field is integrated out one obtains
the initial action.  After obtaining the equation of motion for a general gauge group we specialize to $SU(2)$.
We consider a simple model with two chiral superfields in an $SU(2)$ doublet, and
perform the dualization selecting an Abelian direction inside of $SU(2)$. The results obtained
for a family of Abelian directions apply as well to a set of non-Abelian directions, this
is discussed in Section \ref{Sec:SU2NA}.
In this dual theory where the Lagrange multipliers are the new coordinates,
we obtain the supersymmetric vacuum which, together with the instanton corrections, 
leads to the dual theory.

\subsection{Non Abelian T-duality for a GLSM with global symmetry $G$}
\label{Sec:NATDeom}

In this subsection we start with the general action of a (2,2) GLSM with Abelian gauge group 
$U(1)$, $N$ chiral superfields  $\Phi_{k,i}$  with charges $Q_k$, and a global
symmetry $G$. We gauge the global symmetry by introducing a vector superfield $V$ and Lagrange multipliers.
The vector superfield is integrated to obtain the equations of motion that allow to express the new action in terms of the
Lagrange multipliers coordinates.

We start writing the Lagrangian
\begin{eqnarray}
L_2&=&\int d^4\theta  \left( \sum_k \bar\Phi_{k,i} (e^{2 Q_k V_0+V})_{i j}\Phi_{k,j}+\tr(\Psi\Sigma)+\tr(\bar{\Psi}\bar{\Sigma})-\frac{1}{2 e^2}\bar{\Sigma}_0\Sigma_0\right).\label{nabl}\\
&+&\frac{1}{2} \left(-\int d^2\tilde{\theta} t\Sigma_0+c.c.\right),\nn
\end{eqnarray}
where the Lagrange multipliers $\Psi$ and $\bar \Psi$ are unconstrained superfields in the adjoint representation of the non-Abelian group $G$. When they are integrated out
one obtains the conditions $\Sigma=0$ and $\bar\Sigma=0$. This condition is a pure gauge condition that leads to what we call
{\it the original model}, describing a $U(1)$ gauge theory with chiral superfields coupled to a $U(1)$ vector superfield,
this theory has the Lagrangian:
\begin{eqnarray}
L_1=\int d^4\theta  \left( \sum_k \bar\Phi_{k,i} (e^{2 Q_k V_0})_{i j}\Phi_{k,j}-\frac{1}{2 e^2}\bar{\Sigma}_0\Sigma_0\right)
+\frac{1}{2} \left(-\int d^2\tilde{\theta} t\Sigma_0+c.c.\right).\nn
\end{eqnarray}
Let us now integrate by parts the Lagrange multipliers terms in (\ref{nabl}). We first expand them using the definition of the field strength in (\ref{fieldS}) to get:
\begin{eqnarray}
\int d^4\theta \tr(\Psi\Sigma)&=&\int d^4\theta\tr\left(-\frac{1}{2}(\bar D_+ \Psi) e^{-V} D_{-} e^V\right),\\
\int d^4\theta \tr(\bar\Psi\bar\Sigma)&=&\int d^4\theta\tr\left(-\frac{1}{2}(D_+ \bar\Psi) e^{V} \bar{D}_{-} e^{-V}\right).\nn
\end{eqnarray}
Note that they are complex conjugate of each other. The variation of the exponential of the vector field $\delta e^V$ can  be used to write the variation of the inverse exponential $e^{-V}$ i.e. $\delta e^{-V}$ as:
\begin{equation}
\delta e^{-V}=-e^{-V}(\delta e^V)e^{-V}.
\end{equation}

The variation $\delta e^V$ is not a gauge invariant quantity, a gauge invariant variation of $V$ can be 
defined as \cite{Gates:1983nr} 
\begin{equation}
\Delta V=e^{-V} \delta e^V.
\end{equation}
We can write this variation as $\Delta V=\Delta V_a T_a$, where $T_a$ are the gauge group generators.
The variations with respect to $V$ of the different terms in the Lagrangian (\ref{nabl}) are given next. First, the variation of the kinetic term
for chiral superfields is given by
\begin{eqnarray}
\int d^4\theta \delta \left( \sum_k \bar\Phi_{k,i} (e^{2 Q_k V_0+V})_{i j}\Phi_{k,j}\right)=\int d^4\theta
\sum_k( \bar\Phi_{k,i} (e^{2 Q_k V_0+V})_{i j}(T_a)_{jl}\Phi_{k,l})\Delta V_a,
\end{eqnarray}
Then we compute the variation of the Lagrange multiplier terms:
\begin{eqnarray}
\delta \tr (\bar D_{+} \Psi e^{-V} D_{-} e^V)&=&\tr\left((e^{-V}D_{-} e^V \bar D_{+} \Psi+\bar D_{+} \Psi e^{-V}D_{-} e^V +D_-\bar D_{+} \Psi)\Delta V\right),\nn  \\
\text{ with }\chi&=&e^{-V}D_{-} e^V,\nn
\end{eqnarray}
and the conjugate:
\begin{eqnarray}
\delta \tr (D_{+} \bar\Psi e^{V} \bar D_{-} e^{-V})&=&\tr\left(( -\bar D_{-} e^{-V} D_{+} \bar\Psi e^V+e^{-V} D_{+}\bar \Psi\bar D_{-} e^V +e^{-V}(D_+\bar D_{-} \Psi)e^V)\Delta V)\right).\nn
\end{eqnarray}
Using the previous formulae, the equation of motion obtained for $V$ from $\frac{\delta S}{\delta V}=0$ with the Lagrangian (\ref{nabl}) is given by:
\begin{eqnarray}
-\frac{1}{2}(D_+\bar D_{-} \Psi_b+D_+\bar \Psi_b \times \bar D_{-})\tr(e^V T_ae^{-V}T_b)+(\bar \Phi e^{2 Q V_0}e^V T_a\Phi)\label{eomNA}\\
-\frac{1}{2}(i f_{abc}\tr( e^{-V} D_{-} e^V T_c)\bar{D}_{+} \Psi_b+D_{-}\bar{D}_{+}\Psi_a/2 )=0.\nn
\end{eqnarray}

Now we would like to express $V$ in terms of the dual fields. The gauge symmetry can be used
to fix $\Phi$ and $\bar\Phi$. The vector superfield $V$ is Hermitian, with $V^\dagger=V$, $(e^{\pm V})^\dagger=e^{\pm V}$ and $(\Delta V)^\dagger=\delta e^V e^{-V}=
e^V \Delta V e^{-V}$. This fact helps to simplify the equation of motion. First we notice that $\chi^\dagger=-e^V \bar D_- e^{-V}$ which allows us to write (\ref{eomNA}) as
\begin{eqnarray}
&&\tr((\bar \chi \bar \tau+\bar \tau \bar \chi-\bar D_{-}\bar \tau )T_a)\Delta V^\dagger_a+\tr ((\chi \tau+\tau \chi+D_- \tau )T_a)\Delta V_a\label{teq}\\
&&
+ (\bar \Phi e^{2 Q V_0}e^V T_a\Phi)\Delta V_a=0. \nn
\end{eqnarray}
We make the definition $\bar \tau=D_+\bar \Psi$ and $\tau=\bar D_+ \Psi$.  The Lagrange multipliers terms in the action (\ref{nabl}) can be written as $\tr(\Psi\Sigma)=-\frac{1}{2}\tr(\tau \chi)$ and 
$\tr(\bar \Psi\bar \Sigma)=\frac{1}{2} \tr(\bar \tau \bar \chi)$. Considering a set of commuting generators the equation of motion (\ref{teq}) reduces to the e.o.m. of the Abelian case
given in (\ref{eomAb}).

In the Wess-Zumino(WZ) gauge there is a strong simplification of the equations: 
\begin{eqnarray}
e^{\pm V}=1\pm V+\frac{V^2}{2},\quad e^{-V}D_{\alpha}e^V=D_{\alpha}V-\frac{[V,D_{\alpha} V]}{2}.\label{WZ}
\end{eqnarray}
In this gauge the object $\chi=e^{-V}D_- e^V$ can be expressed
as $\chi=\chi_aT_a$, this may be possible in other gauges as well.
For $SU(2)$ we have $\chi=\chi_a\sigma_a$ at any gauge.
Then (\ref{teq}) simplifies to
\begin{eqnarray}
&&D_+\bar D_{-} \bar \Psi _a\Delta V^\dagger_a+
D_- \bar D_+ \Psi_a\Delta V_a+(\bar \Phi e^{2 Q V_0}e^V T_a\Phi)\Delta V_a=0.\label{teq20}
\end{eqnarray}

The previous equation contains the variation $\Delta V_a$ and its conjugate. Now let us argue that they 
coincide. As we noticed previously $(\Delta V)^\dagger=e^V \Delta V e^{-V}$, this implies
$(\Delta V_a)^\dagger T_a=(e^V T_a e^{-V} )\Delta V_a$, which for
$\tr T_a \neq 0$  gives $\Delta V_a=(\Delta V_a)^\dagger$.
However if $\tr T_a=0$ the relation gives 
$\Re(\Delta V_a)(T_a-e^V T_a e^{-V})+i \Im(\Delta V_a)(T_a+e^V T_a e^{-V})=0$.
The mentioned set of equations has solutions $\Im(\Delta V_a)=0$ and 
$\Delta V_a T_a=e^V \Delta V_aT_a e^{-V}$.  This last condition 
implies $[V,\Delta V]=0$ which is a condition that arises also from the Wess
Zumino gauge relations (\ref{WZ})  if $\Delta V^\dagger=e^V \Delta V e^{-V}$ holds.
Therefore we will consider that the relation $\Delta V_a=(\Delta V_a)^\dagger$ holds,
and look at the equation of motion (\ref{teq}) under this consideration.

\subsection{GLSM with an $SU(2)$ symmetry}
\label{SU2glsm}

We now consider a gauge linear sigma model with global $SU(2)$ symmetry that can be gauged.
This is the explicit example that we explore in detail in this work. We consider the case
of a $U(1)$ gauge symmetry and two chiral superfields equally charged under it. We apply
our procedure of gauging the global symmetry of the theory which in this case is an $SU(2)$
symmetry. Integrating the vector superfield via its e.o.m.  leads to the dual model.

The Lagrangian which includes as limits the two dual theories is given by
\begin{eqnarray}
L_{model}&=&\int d^4\theta  \left(  \bar\Phi_i (e^{2 Q V_0+V})_{i j}\Phi_{j}+\tr(\Psi\Sigma)+\tr(\bar{\Psi}\bar{\Sigma})-\frac{1}{2 e^2}\bar{\Sigma}_0\Sigma_0 \right)\label{nablsu2}\\
&-&\frac{1}{2} \left(\int d^2\tilde{\theta} t\Sigma_0+c.c.\right),\nn
\end{eqnarray} 
where by $V_0$ and $\Sigma_0$ we denote the vector superfield and field strength of the $U(1)$ gauge group of the GLSM, $\Phi_i$ and $\bar \Phi_i$ with indices $i,j=1,2$ denote the two chiral and  anti-chiral superfields, $V$
is the gauge field of the $SU(2)$ gauged symmetry with field strength $\Sigma$ and $\Psi$ is the Lagrange multiplier, an unconstrained superfield.
The  superfields $\Phi$ and $\bar \Phi$ are doublets under the $SU(2)$  global symmetry. 
The original model is obtained by integrating  out the Lagrange multiplier $\Psi$ obtaining a pure gauge configuration $\Sigma=0$. This
defines the $\mathbb{CP}^1$ linear sigma model. The scalar potential in this model is given by
\begin{eqnarray}
U=\frac{e^2 Q^2}{2}(|\phi_1|^2+|\phi_2|^2-2r)^2+2|\sigma_0|^2Q^2(|\phi_1|^2+|\phi_2|^2).
\end{eqnarray}
The vacuum manifold for $r>0$ is given by  $\mathbb{CP}^1\equiv \{ |\phi_1|^2+|\phi_2|^2=2r\}/U(1)$ and $\sigma_0=0$. 
The mirror of  the $\mathbb{CP}^1$ model is obtained by dualizing with Abelian T-duality along two directions i.e.
employing the two rotation symmetries of the two fields $\Phi_1$ and $\Phi_2$ is the $A_1$ Toda field theory \cite{Hori:2000kt}
which will be compared with our dual models in Section \ref{Sec:Analysis}. Now we want to use the global $SU(2)$ symmetry to implement non-Abelian T-duality
in this model.

For this model the effective superpotential in the original GLSM for the gauge field $\Sigma_0$ (\ref{effW}) reads \cite{Witten:1993yc}
\begin{eqnarray}
W_{eff}(\Sigma_0)= -2 Q \Sigma_0 \ln\left(\frac{Q\Sigma_0}{\mu}\right)+2 Q \Sigma_0-t\Sigma_0.\label{effW1}
\end{eqnarray}

Due to the coincident charges the theory has one chiral and one antichiral $SU(2) $ doublets
$(\Phi_1,\Phi_2)$ and $(\bar \Phi_1,\bar \Phi_2)$ respectively.
Let us write the e.o.m. of the gauged vector superfield $V $(\ref{teq}) as
\begin{eqnarray}
\tr((\bar \chi \bar \tau+\bar \tau \bar \chi-\bar D_{-}\bar \tau )T_a)+
\tr ((\chi \tau+\tau \chi+D_- \tau )T_a)+
(\bar \Phi e^{2 Q V_0}e^V T_a\Phi)=0 \label{teq2}, 
\end{eqnarray}
Let us define:
\begin{eqnarray}
\bar X_a=\tr((\bar \chi \bar \tau+\bar \tau \bar \chi-\bar D_{-}\bar \tau )T_a), \,  \, \, X_a=\tr ((\chi \tau+\tau \chi+D_- \tau )T_a).\nn
\end{eqnarray}
Employing (\ref{teq2}) it is possible to express $ (\bar \Phi e^{2 Q V_0}e^V \Phi)$ in terms of  $V$, thus eliminating
3 of the 4 products $\bar \Phi_i \Phi_j$, and leaving one. Applying the property $\det e^V=1$ one obtains
\begin{eqnarray}
e^{2 Q V_0}\bar \Phi_i e^V_{ij}\Phi_j&=&-e^{2 Q V_0}\frac{2 \bar\Phi_1 \Phi_2}{e^V_{21}} - \frac{e^V_{22} (X_1+ i X_2+c.c.)}{e^V_{21}}  - X_3+c.c.\label{sol1}\\
X_a&=&\tr ((\chi \tau+\tau \chi+D_- \tau )\sigma_a)=D_- \bar D_+ \Psi_a .\nn
\end{eqnarray}
The superfields $X_a$ gets simplified\footnote{Taking into account that $\chi=\chi_a\sigma_a$, for $SU(2)$ this happens
in any gauge, we thus obtain
$ \tr ((\chi \tau+\tau)\sigma_a)=$$\tr ((\chi_b \tau_c+\chi_c \tau_b )\sigma_b \sigma_c\sigma_a)$
$= \tr ((\chi_b \tau_c+\chi_c \tau_b )\epsilon_{bca}2i)=0$. For other Lie algebras  there is also a symmetric
part $d_{abc}$ in the product $T_aT_b$, check (\ref{TaTb}), this makes the mentioned term different from zero.} in $SU(2)$.
It is  possible to use the gauge freedom in the gauged model to set $\bar\Phi_1 \Phi_2=0$, by setting
one of those components to zero. Substituting (\ref{sol1}) in (\ref{nabl}) we arrive to the dual model, which we analyze
in the following. Let us look at the reduction to the Abelian case considering $\chi=e^{-V}D_- e^V=D_- V$
and only one generator $\sigma_a=1$, one gets:
\begin{eqnarray}
\tr ((\chi \tau+\tau \chi+D_- \tau )\sigma_a)+c.c.&=&\tr ((D_- V \tau+\tau D_- V +D_- \bar D_+ \Psi)),\\
&=&D_- \bar D_+ \Psi+\bar D_- D_+ \bar \Psi.\nn\\
&=&\Lambda+\bar \Lambda .\nn
\end{eqnarray}
This equation coincides with the result for the e.o.m. of the Abelian duality given in (\ref{eomAb}).

Let us work the $V$ equation of motion (\ref{teq2}) to solve
for $V$ vs. the other fields. Simplifying (\ref{teq2}) with the condition $\tr(\chi \tau+\tau\chi)=0$  one gets
\begin{eqnarray}
(\bar \Phi e^{2 Q V_0}e^V \sigma_a\Phi)&=&\bar D_{-}D_+ \bar \Psi _a+
\bar D_+D_-  \Psi_a=\bar X_a+X_a.\label{teq3}
\end{eqnarray}
These constitute three equations of motion, one for each generator of the group algebra. Note that on (\ref{teq3})
the fundamental fields in the duality become $X_a$ and $\bar X_a$ which are twisted chiral
and twisted anti-chiral superfields. This is the case for $SU(2)$, but in more general cases
the fundamental fields appear to be the semi-chiral superfields $\bar D_+ \Psi_a$,
recall the e.o.m. (\ref{teq}). This needs to be explored further for other gauged groups.

The three equations of motion, one for each group generator
can be employed to express two components of
$e^V$ in terms of the other two. \footnote{There is an extra relation coming from conjugating the previous expression
$\bar \Phi_i e^{2 Q V_0}[e^V, \sigma_a]_{ij}\Phi_j=0$. This could be achieved by setting $[e^V, \sigma_a]=0$.}  We write
the e.o.m. (\ref{teq3}) for the generators $a=1,2$ as
\begin{eqnarray}
e^V_{12}&=&-e^V_{22} \frac{\bar\Phi_2}{\bar\Phi_1} +\frac{X_1+\bar X_1- i (X_2+\bar X_2)}{2 e^{2 Q V_0} \Phi_1 \bar\Phi_1},\label{VvsPhi}\\
e^V_{21}&=&-e^V_{11} \frac{\bar\Phi_1}{ \bar\Phi_2} + \frac{X_1+\bar X_1+ i (X_2+\bar X_2)}{2 e^{2 Q V_0} \Phi_2 \bar\Phi_2},\nn
 \end{eqnarray} 
where $e^V_{ij}$ denotes the $(i,j)$ matrix element of $e^V$. Employing these relations it is possible to eliminate the $V$ dependence in the kinetic term of (\ref{nablsu2}) to get:
\begin{eqnarray}  
e^{2 Q V_0} \bar \Phi_{i} e^V_{ij} \Phi_j= \frac{-((\Phi_1^2 (X_1 + \bar X_1+ i( X_2  + \bar X_2)) + 
\Phi_2^2 (X_1+\bar X_1 - i (X_2 +  \bar X_2))}{(2 \Phi_1 \Phi_2)}
  \end{eqnarray} 
The third constraint coming from (\ref{teq3}) evaluated for $\sigma_3$ reads:
\begin{eqnarray}  
&&\Phi_2^2  (X_1+\bar X_1- i (X_2+\bar X_2)) - \Phi_1^2  (X_1+\bar X_1+i (X_2+\bar X_2))+\label{eqPhiX1} \\
&& 2 \Phi_1 \Phi_2 (X_3+\bar X_3)=0,\nn
\end{eqnarray} 
 which allows to express the chiral superfields of the model vs. the twisted superfields
 of the dual model. One can divide (\ref{eqPhiX1}) by $\Phi_1^2$ to obtain
 \begin{eqnarray}  
&&(\Phi_2/\Phi_1)^2  (X_1+\bar X_1- i (X_2+\bar X_2)) -  (X_1+\bar X_1+i (X_2+\bar X_2))\label{eqPhi}\\
&&+ 2 (\Phi_2/\Phi_1) (X_3+\bar X_3)=0,\nn
\end{eqnarray} 
The solution reads:
 \begin{eqnarray}  
F=(\Phi_2/\Phi_1)=\frac{-(X_3+\bar X_3)\pm \sqrt{(X_3+\bar X_3)^2+((X_1+\bar X_1)^2+(X_2+\bar X_2)^2)})}{(X_1+\bar X_1- i (X_2+\bar X_2)) }.\label{P1P2}
\end{eqnarray} 
Let us choose the solution with $+$. This solution can be employed to express the kinetic term of the Lagrangian in terms of the dual coordinates $X_1,X_2,X_3$:
 \begin{eqnarray}  
e^{2 Q V_0} \bar \Phi_{i} e^V_{ij} \Phi_j= \sqrt{( X_1  + \bar X_1)^2+ (X_2+\bar X_2)^2+ (X_3 +  \bar X_3)^2}.\label{kinDual}
  \end{eqnarray} 
From the chosen solution with plus on (\ref{P1P2}),  one can use the gauge freedom to set $F=1$, i.e.,
giving $\Phi_1=\Phi_2$, which implies a constraint between the real parts of the twisted chiral superfields 
\begin{eqnarray}
\Phi_1=\Phi_2 \rightarrow X_2+\bar X_2=X_3+\bar X_3=0.\label{gauge1}
\end{eqnarray}
One could also fix 
$\Phi_2=i\Phi_1$, then the gauge implies $X_1+\bar X_1=X_3+\bar X_3=0$.  This will be particularly useful at the
time of analyzing the dual model geometry.

\subsection{Dualization  along $\sigma_1\in SU(2)$, $V=|V|\sigma_1$}
\label{Subsec:T1}
In this subsection we would like to proceed with a simple case, that is to take an Abelian direction inside the duality group along the generator $\sigma_1$. In this scheme we perform the duality and analyze the obtained dual theory.

We start writing the vector superfield $V$ aligned with the generator $\sigma_1$, i.e. $V=V_1\sigma_1$, for the exponential of $V$ that is:
\begin{eqnarray}
e^{V}=e^{V_1 \sigma_1}= \left( \begin{array}{cc}
    \cosh(V_1) & \sinh(V_1) \\ 
   \sinh(V_1) & \cosh(V_1) \\ 
  \end{array}\right) ,\,  |V|=V_1,\, n_1=1.\label{eV1eq}
\end{eqnarray}  
In Appendix \ref{SU2formulas} we present relevant formulas for $SU(2)$ vector superfields and its equations of motion,
in particular we write the expression that leads to (\ref{eV1eq}). Using the equations of motion from the generators $\sigma_1$ and $\sigma_2$ in  (\ref{teq3}) one obtains
\begin{eqnarray}
\sinh(V_1) + (2 g_p \phi_1 \bar\phi_2 \cosh(V_1) - X_1 - \bar X_1 + i( X_2+ \bar X_2))/(
 2 g_p |\phi_1|^2)=0,\\ 
 \sinh(V_1) + (2 g_p \bar\phi_1 \phi_2 \cosh(V_1) - X_1 - \bar X_1 - i( X_2 + \bar X_2))/(
 2 g_p |\phi_2 |^2)=0.
 \end{eqnarray}
Those equations can be solved to obtain
\begin{eqnarray}
e^{V_1}&=&\frac{\phi_1 (X_1 + \bar X_1 + i( X_2 +  \bar X_2)) + 
\phi_2 (X_1+ \bar X_1- i (X_2+ \bar X_2))}{2 e^{2 Q V_0} \phi_1 \phi_2 (\bar\phi_1 + \bar \phi_2)},\label{eV1}\\
&=&\frac{ (X_1 + \bar X_1 + i( X_2 +  \bar X_2))}{2 e^{2 Q V_0}  \phi_2 (\bar\phi_1 + \bar \phi_2)}
+\frac{ (X_1+ \bar X_1- i (X_2+ \bar X_2))}{2 e^{2 Q V_0} \phi_1 (\bar\phi_1 + \bar \phi_2)}.\nn
\end{eqnarray}
From it we get
\begin{eqnarray}
V_1=\ln\left((X_1 + \bar X_1)+ i( X_2 +  \bar X_2)\frac{\bar F-1}{\bar F+1}\right)-\ln (2 |\Phi_1|^2)-2 Q V_0.\label{V1eq}
\end{eqnarray}
The fact that $V_1$ is a real vector superfield implies relations among the twisted chiral and the twisted anti-chiral superfields such that
the argument of the logarithm is real.
Let us use the relations:
\begin{eqnarray}
\phi_1 (\bar\Phi_1 + \bar \Phi_2)&=&|\Phi_1|^2(1 +\bar F),\label{V10}\\
\phi_2 (\bar\Phi_1 + \bar \Phi_2)&=&|\Phi_2|^2(1 +1/\bar F),\\
|\Phi_1|^2&=&|\Phi_2|^2,
\end{eqnarray}
with $F=\Phi_2/\Phi_1$ given by (\ref{P1P2}). One can show that $V_1$ satisfiying (\ref{eV1}) fulfills $(e^{V_1})^{\dagger}=e^{V_1}$. One also has
 \begin{eqnarray}
e^{-V_1}&=&\frac{\phi_1 (X_1 + \bar X_1 + i( X_2 +  \bar X_2)) - 
\phi_2 (X_1+ \bar X_1- i (X_2+ \bar X_2))}{2 e^{2 Q V_0} \phi_1 \phi_2 (\bar\phi_1 - \bar \phi_2)},\label{V10Inv}\\
&=&\frac{ (X_1 + \bar X_1 + i( X_2 +  \bar X_2))}{2 e^{2 Q V_0}  \phi_2 (\bar\phi_1 - \bar \phi_2)}
-\frac{ (X_1+ \bar X_1- i (X_2+ \bar X_2))}{2 e^{2 Q V_0} \phi_1 (\bar\phi_1 - \bar \phi_2)},\nn
\end{eqnarray}
 with $(e^{-V_1})^{\dagger}=e^{-V_1}$.  As a check, it is possible to verify that $e^{-V_1}e^{V_1}=1$ using the equations (\ref{V10}) and (\ref{V10Inv}).
There are three contributions to $\int d\theta^4 (\tr \Psi \Sigma+\tr \bar \Psi \bar \Sigma)$ coming from the three terms
in (\ref{V1eq}), those are:
\begin{eqnarray}
 \int d^4\theta (X_1 V_1+\bar X_1 V_1)&\supset& \int d^4\theta  X_1\ln((X_1+\bar X_1)+i  (X_2+\bar X_2) \frac{(\bar F-1)}{(\bar F+1)})+c.c,\label{V1eq1}
\end{eqnarray}
from the first term, also:
\begin{eqnarray}
 \int d^4\theta (X_1 V_1+\bar X_1 V_1)&\supset&  \int d^4\theta (X_1 +\bar X_1)\ln (2 \bar \Phi_1\Phi_1),\\
&=&0.\nn
\end{eqnarray}
coming from the second term, and:
\begin{eqnarray}
\int d^4\theta (X_1 V_1+\bar X_1 V_1)&\supset&- \int d^4\theta(D_-\bar D_+\Psi_1 2 Q V_0+ D_+\bar D_-\bar \Psi_1 2 Q V_0),\label{V1eq2}\\
 &=&-\frac{2 Q}{2} \int d\theta^+ d\bar \theta^-D_-\bar D_+\Psi_1 \Sigma_0-\frac{2 Q}{2}\int d\theta^- d\bar \theta^+ D_+\bar D_-\bar \Psi_1 \bar \Sigma_0,\nn\\
 &=&-\frac{2 Q}{2}\int d\theta^+ d\bar \theta^-X_1 \Sigma_0-\frac{2 Q}{2}\int d\theta^- d\bar \theta^+\bar X_1 \bar \Sigma_0,\nn
\end{eqnarray} 
coming from the third term. This last contribution is a duality generated twisted superpotential. Thus, in this gauge we are able to eliminate
the original fields and the gauged field in terms of the Lagrange multipliers which become the dynamical fields. 

 The dual Lagrangian can be obtained starting with (\ref{nablsu2}) and using all previous relations (\ref{kinDual}),(\ref{V1eq1}) and (\ref{V1eq2}) to get:
 \begin{eqnarray}
L_{dual}&=&\int d^4\theta \sqrt{( X_1  + \bar X_1)^2+ (X_2+\bar X_2)^2+ (X_3 +  \bar X_3)^2}+\label{Ldual1}\\
&+&\int d^4\theta  X_1 \ln((X_1+\bar X_1)+i  (X_2+\bar X_2) (\bar F-1)/(\bar F+1))+c.c.\nn\\
&+&\frac{2 Q}{2}\int d\bar \theta^-d\theta^+\left(X_1-\frac{t}{2Q}\right) \Sigma_0+\frac{2 Q}{2}\int d\bar \theta^+d\theta^-\left(\bar X_1-\frac{\bar t}{2Q}\right) \bar \Sigma_0- \int d^4\theta\frac{1}{2 e^2}\bar{\Sigma}_0\Sigma_0.\nn
\end{eqnarray}
 
 The last term is a twisted superpotential generated by the duality. One can make an Ansatz for the instanton corrections
on the dual theory, compute the effective superpotential for $\Sigma_0$ and compare it with the one loop calculation done for
that quantity in the original theory \cite{Witten:1993yc}. Since we are dualizing an Abelian direction inside the
gauged group $SU(2)$ we will take the Ansatz for the instanton corrections to be the same as the Hori and Vafa \cite{Hori:2000kt} $\Delta W=2\mu e^{-X_1}$ giving the total twisted superpotential
\begin{eqnarray}
\widetilde W=2 Q X_1 \Sigma_0+2\mu e^{-X_1}-t\Sigma_0.\label{W1}
\end{eqnarray}
By integrating out $X_1$ one gets $X_1=-\ln (\frac{Q\Sigma_0}{\mu})$ and by plugging this into (\ref{W1})
we obtain
\begin{eqnarray}
\widetilde W_{eff}(\Sigma_0)'=-2 Q\ln \left(\frac{Q\Sigma_0}{\mu}\right)+2 Q \Sigma_0-t\Sigma_0.\label{effW2}
\end{eqnarray}
This last expression coincides with (\ref{effW1}), the effective superpotential for $\Sigma_0$ on the original theory. This gives evidence for the equivalence of both theories. Next, we analyze the supersymmetric vacua of the dual theory.

\paragraph{Scalar potential of the dual model.}

Let us look at the scalar potential for the twisted field strength of the GLSM and the twisted chiral superfields arising in the dual theory.
The vanishing of the new potential is what will give the susy locus, i.e., the target space geometry of the dual theory. 

Using the expansions for the twisted superfields given in (\ref{twistedF}) one can express the twisted superpotential in terms of the scalar components of the twisted superfields. All the terms contributing to (\ref{Ldual1}) read
\begin{eqnarray}
\frac{2 Q}{2}\int d\bar \theta^-d\theta^+X_1 \Sigma_0&=&2 Q( i(\chi_{-,1} \bar\lambda_{+,0}+\bar  \chi_{+,1} \lambda_{-,0})+\sigma_0 G_1+ x_1(D-i v_{03})),\\
\frac{2 Q}{2}\int d\bar \theta^+d\theta^-\bar X_1 \bar \Sigma_0&=&2 Q(i(\bar \chi_{-,1} \lambda_{+,0}+\chi_{+,1} \bar\lambda_{-,0})+\bar \sigma_0 \bar G_1+\bar x_1(D+i \bar v_{03})), \nn\\
-\frac{1}{2 e^2}\int d \theta^4\bar \Sigma_0\Sigma_0&=&\frac{1}{2 e^2} |D-i v_{03}|^2, \nn\\
 -2 Q\frac{t}{4Q}\int d\bar \theta^-d\theta^+ \Sigma_0&=&-Q t (D-i v_{03}),\nn \\
  -2 Q\frac{\bar t}{4Q}\int d\bar \theta^+d\theta^- \bar\Sigma_0&=&-Q \bar t (D+i v_{03}).\nn
\end{eqnarray}

The kinetic part for the scalar components of the twisted chiral superfields is given by 
\cite{rocek84,wessbagger}
\begin{eqnarray}
S_{scalar}=-\frac{1}{2}\int d^2 x( -K_{i\bar j} \partial_a x_i \partial^a \bar x_j+K_{i\bar j} G_i  \bar G_j). \label{ScalarT}
\end{eqnarray}
with $i,j=1,2,3$.\footnote{Notice that in our notation we use the index $\bar j$ to denote the derivative of
$K$ with respect to $\bar x_j$, but to denote the conjugate superfield $\bar X_j$ and its components we use the index $j$.}  In Appendix \ref{twistedExpansion} we write generic expressions to compute the Lagrangian contribution
of twisted chiral superfields for a given K\"ahler potential and twisted superpotential.  In particular, we derive the second term in (\ref{ScalarT}). Collecting all the contributions to the scalar potential depending on the dynamical fields $\sigma_0, x_a$ and the auxiliary fields
$D$ and $G_i$ one gets:
\begin{eqnarray}
U=-4 Q Re(\sigma_0 G_1)-4 Q Re((x_1-t/8)D)-\frac{1}{2 e^2} D^2+K_{i\bar j} G_i  \bar G_{j}.\label{U10}
\end{eqnarray}
It is necessary to compute the equations of motion for the auxiliary fields. Solving for
the auxiliary field $D$,\footnote{Note that we have omitted fermionic terms in the e.o.m. for auxiliary fields because we are interested
here just in the scalar potential.} yields $D=-2 e^2 Q(x_1+\bar x_1-(t+\bar t)/(2Q))$ and 
one gets the scalar potential
\begin{eqnarray}
U=-4 Q Re(\sigma_0 G_1)+4 Q^2 e^2 (x_1+\bar x_1-(t+\bar t) /(2Q))^2+K_{i\bar j} G_i  \bar G_j.
\end{eqnarray}
The above expression for the scalar potential suggests that we use the equation of motion for $G_1$ in order to eliminate  the auxiliary fields completely.
One obtains $K_{2\bar j}  \bar G_j=K_{3\bar j}  \bar G_j=0$ and $K_{1\bar j}  \bar G_j=2 Q \sigma_0$.
The previous system is solved by the condition $\sigma_0=0$.
Thus we obtain the potential:
\begin{eqnarray}
U=2 Q^2 e^2 (x_1+\bar x_1-(t+\bar t)/(2Q))^2,
\end{eqnarray}
whose zero locus is $\Re x_1=\Re t/(2Q)=r/(2Q)$ which represents the susy vacuum in the dual theory.

However, if one looks at the Higgs branch in the NLSM limit the field $\Sigma$ can be integrated out \cite{Hori:2000kt},
which in this description means to integrate $v_{03}$. To obtain an effective scalar potential one starts from the
scalar potential plus the interaction terms with $v_{03}$:
\begin{eqnarray}
U_{eff}=-\frac{1}{2 e^2}( v_{03}^2+v_{03}2 i e^2 Q (t - \bar t - 2Q n_a x_a + 2Q n_a \bar x_a))+2 Q^2 e^2 (x_1+\bar x_1-(t+\bar t) /(2Q))^2.\nn
\end{eqnarray}
Integrating with respect to $v_{03}$ one obtains $v_{03}=- i e^2 Q (t - \bar t - 2Q n_1( x_1-\bar x_1))$ 
which leads to the scalar potential
\begin{eqnarray}
U_{eff}=2 Q^2 e^2 \lvert x_1-\frac{t}{2Q} \rvert^2,
\end{eqnarray}
whose zero locus $|x_1-t/(2Q)|^2=0$ represents the dual manifold. Adding the instanton contributions leads us to 
the dual model in the IR limit. It is a very simple model given by $X_1=t/(2Q)$ and constant twisted
superpotential coming from (\ref{W1}):  $\widetilde W=2\mu e^{-t/(2Q)}$. Next, we are going to generalize
the study for a family of Abelian directions inside the $SU(2)$ gauged group.

\subsection{Dualization along $n_a\sigma_a \in SU(2)$,  $V=|V|n_a \sigma_a$}
\label{Subsec:Ta}

In this subsection we perform a family of Abelian dualities along the direction of a linear combination of the generators,
leading to $V=|V|n_a \sigma_a$. We obtain the dual sigma model, in particular the scalar potential from which we analyze
the susy vacua. The scalar potential of the  dual theory also describes truly non-Abelian models. There is a Higgs branch which leads to the integration of the original
$U(1)$ field strength and to a dual theory. We make an Ansatz for the instanton contributions
to the twisted superpotential which leads to a correct effective potential for $\Sigma_0$, and after the integration
on $\Sigma_0$ leads to a dual Abelian theory.

In this case the Lagrange multiplier term is given by $\int d^4\theta\tr (\Psi \Sigma)=\int d^4\theta \frac{1}{2}X_a \hat n_a |V|$. From (\ref{teq3}) we have:
\begin{eqnarray}
|V|=\ln(\mathcal{K}(X_a,\bar X_a,\hat n_a))-2 Q V_0-\ln 2 |\Phi_1|^2.
\end{eqnarray}
As before the fact that $|V|$ is a real vector superfield implies that $\mathcal{K}(X_a,\bar X_a,\hat n_a)$ is as well real, this gives a 
relation among the twisted chiral and the twisted anti-chiral superfields. The expression for $\mathcal{K}$ is given
in equation (\ref{KLog}) of Appendix \ref{SU2formulas}. The gauge fixing condition  (\ref{gauge1}) $X_2+\bar X_2=X_3+\bar X_3=0$, guarantees $\Im(\mathcal{K}(X_a,\bar X_a,\hat n_a))=0$.
In Appendix \ref{SU2formulas} we gathered useful formulae for $SU(2)$ algebra and the vector superfields
taking values in this algebra.

Note that, in principle, the $\hat n_a$ depend on the superspace coordinates
with restriction $\hat n_a^\dagger=\hat n_a$. Later in Section \ref{Sec:SU2NA} we will see that if $\hat n_a$ is a real
superfield with the restriction $D_{-}\hat n_a=0$ one can also integrate the equations to have the dual Lagrangian
in terms of twisted chiral superfields. 
The dual Lagrangian for $\hat n_a$ constant may be written as
 \begin{eqnarray}
L_{dual}&=&\int d^4\theta \sqrt{( X_1  + \bar X_1)^2+ (X_2+\bar X_2)^2+ (X_3 +  \bar X_3)^2}+\label{Sdual1}\\
&+& \int d^4\theta  X_a \hat n_a \ln(\mathcal{K}(X_i,\bar X_i,n_j))+c.c.- \int d^4\theta\frac{1}{2 e^2}\bar{\Sigma}_0\Sigma_0\nn\\
&+&\frac{2 Q}{2}\int d\bar \theta^-d\theta^+\left(X_a n_a-\frac{t}{2Q}\right) \Sigma_0+\frac{2 Q}{2}\int d\bar \theta^+d\theta^-\left(\bar X_a n_a-\frac{\bar t}{2Q}\right) \bar \Sigma_0.\nn
\end{eqnarray}
The action (\ref{Sdual1}) is invariant under the transformation $X_a\rightarrow X_a+\frac{2\pi i
 k_a}{2 Q n_a},\, k_a\in \mathbb{Z}$, taking into account that the $\theta$ angle has quantum symmetry $t\rightarrow t+2 \pi i$.

To compute the scalar potential let us first consider the case of constant $n_a$\footnote{Modifications to this condition should give a truly non-Abelian direction inside the non-Abelian group that will be studied in Section \ref{Sec:SU2NA}. }. Thus we obtain
\begin{eqnarray}
\int d\bar \theta^-d\theta^+X_a n_a \Sigma_0&=&2 n_a i(\chi_{-,a} \bar\lambda_{+,0}+\bar  \chi_{+,a} \lambda_{-,0})+2 \sigma_0 G_a n_a+2 x_a n_a(D-i v_{03}),\label{abinside}\\
\int d\bar \theta^+d\theta^-\bar X_a n_a \bar \Sigma_0&=&2 n_ai(\bar \chi_{-,a} \lambda_{+,0}+\chi_{+,a} \bar\lambda_{-,0})+2 \bar \sigma_0 \bar G_a n_a+2 \bar x_a n_a(\bar D+i \bar v_{03}).
\end{eqnarray}
The scalar potential in terms of the dynamical and auxiliary fields in this case is given by
\begin{eqnarray}
U=-2 Q(\sigma_0 G_a+\bar \sigma_0 \bar G_a) n_a-4 Q \Re((x_a n_a-t/(2Q))D)-\frac{1}{2 e^2} D^2+K_{i\bar j} G_i  \bar G_j.\nn
\end{eqnarray}
Integrating as usual the auxiliary field $D$ one gets\footnote{ As discussed in \cite{Hori:2000kt} in the Higgs branch one can also integrate $\Sigma$; in this scheme it is related to fixing  $\sigma_0$ at its vanishing vev and to integrate $v_{03}$ which  leads to the scalar potential.}
\begin{eqnarray}
\label{doublet}
U=-2 Q(\sigma_0 G_a+\bar \sigma_0 \bar G_a) n_a+K_{a\bar b} G_a \bar G_b+2 Q^2 e^2 \left(\sum_a (x_a+\bar x_a) n_a-(t+\bar t)/(2Q)\right)^2.
\end{eqnarray}
Integrating also the auxiliary field $G_a$ we obtain $K_{a\bar b}\bar G_{\bar b}=2 Q \sigma_0 n_a$. 
This only lets a contribution to the scalar potential given by $2 Q \bar \sigma_0 \bar G_a n_a\subset U$.
Now the constraints are solved to give
\begin{eqnarray}
U&=&-2 Q \bar \sigma_0 \bar G_a n_a+2 Q^2 e^2 \left(\sum_a  (x_a+\bar x_a) n_a-(t+\bar t)/(2Q)\right)^2 \label{Udual}\\
&=&8 Q^2|\sigma_0|^2\mathcal{O}(x_i,\bar x_i,n_j)+2 Q^2 e^2 \left(\sum_a  (x_a+\bar x_a) n_a-(t+\bar t)/(2Q)\right)^2.\nn
\end{eqnarray}
 The function $\mathcal{O}(x_i,\bar x_i,n_j)$ depends on the components of the K\"ahler metric. A susy vacuum
appears at $\sigma_0=0$ and $\Re(\sum_a (x_a+\bar x_a) n_a)=\Re t/(2Q)=r/(2Q)$. Recall that the condition $n_a n_a=1$ holds, see Appendix \ref{SU2formulas}, below equation (\ref{VaApp}).  


In the gauge (\ref{gauge1}) the K\"ahler potential depends only on $X_1+\bar X_1$,
this is $K= \sqrt{(X_1+\bar X_1)^2} + n_1 (X_1+\bar X_1) \ln\left(\sqrt{(X_1+\bar X_1)^2}/2\right)$. Integrating
out the auxiliary fields 
\begin{eqnarray}
0=2 Q \sigma_0 n_2, \quad 0=2 Q \sigma_0 n_3, \quad K_{1\bar 1}\bar G_{\bar 1}=2 Q \sigma_0 n_1.
\end{eqnarray}
This is solved as $n_2=n_3=0$ or $\sigma_0=0$. This makes $\mathcal{O}(x_i,\bar x_i,n_j)=-n_1^2/K_{1\bar 1}=- n_1 (X_1+ \bar X_1)$ and the potential is given by
\begin{eqnarray}
U&=&-8 Q^2|\sigma_0|^2\frac{n_1^2}{K_{1\bar 1}}+2 Q^2 e^2 \left(\sum_a  (x_a+\bar x_a) n_a-(t+\bar t)/(2Q)\right)^2,\label{vacioCoul}\\
&=&-8 Q^2|\sigma_0|^2n_1 (x_1 + \bar x_1)+2 Q^2 e^2 \left(\sum_a  (x_a+\bar x_a) n_a-(t+\bar t)/(2Q)\right)^2.\nn
\end{eqnarray}
A vacuum appears at $\Re(x_1)=\frac{\Re(t)}{2Q}$, $|\sigma_0|=0$ which is located at
$U=0$ and is a stable minimum at $\Re(t)<0$. There is no minimum on the Coulomb branch.

Let us now make a similar analysis as the one performed in the previous Section.  On the Higgs branch one can go to the
IR limit by integrating out the $U(1)$ field strength $\Sigma_0$. As $\sigma_0=0$ this is equivalent to integrate $v_{03}$ considering
it as the fundamental field. We start with the effective potential including (\ref{vacioCoul}) and the interactions with
$v_{03}$ to have
\begin{eqnarray}
U_{eff}&=&-\frac{v_{03}^2}{2 e^2}-v_{03} iQ (t - \bar t - 2Q \sum_a n_a(x_a + \bar x_a)))+\label{UeffTa0}\\
&+&2 Q^2 e^2 \left(\sum_a  (x_a+\bar x_a) n_a-(t+\bar t)/(2Q)\right)^2.\nn
\end{eqnarray}
In the last step we have integrated $v_{03}$ to obtain an effective potential in IR theory:
\begin{eqnarray}
U_{eff}=2 Q^2 e^2 |\sum_a x_a n_a-t/(2Q)|^2.\label{UeffTa}
\end{eqnarray} 
The dual manifold of this theory is at the locus $\sum_a x_a n_a-\frac{t}{2Q}=0$. This is a hyperplane in $\mathbb{C}^3$.
But there are additional restrictions from gauge fixing and from the instanton corrections in the twisted superpotential.

We have obtained a family of dual models  by gauging a $U(1)$ symmetry inside the global $SU(2)$ of the original model (obtained from (\ref{nablsu2}) after integrating the Lagrange multiplier).
Adding an Ansatz for the expected instanton contributions of the dual models leads to the twisted superpotential
\begin{eqnarray}
\widetilde W=2 Q X_a n_a \Sigma_0-t\Sigma_0+2\mu e^{-X_a n_a}.\label{Wna}
\end{eqnarray}
By integrating out $X_a$ one gets $X_a n_a=-\ln (\frac{Q\Sigma_0}{\mu})$ and by plugging this into (\ref{Wna})
we obtain again $\widetilde W_{eff}(\Sigma_0)'=-2 Q\ln \left(\frac{Q\Sigma_0}{\mu}\right)+2 Q \Sigma_0-t\Sigma_0$
which is exactly the effective superpotential for the $U(1)$ field strength of the GLSM(\ref{effW2}).

The vacuum space comes from the following restrictions on the coordinates. Going to the Higgs branch, i.e.,  $\sigma_0=0$ in the IR limit one can integrate the gauge $U(1)$ field
$\Sigma_0$ \cite{Hori:2000kt} in the twisted superpotential to obtain the condition $X_an_a=\frac{t}{2 Q}$. This gives a constant
twisted superpotential $\widetilde W=e^{-t/(2 Q)}$, with scalar potential $U=0$. Recall moreover
that one can fix the gauge symmetry to obtain $\Phi_1=\Phi_2$ and this will imply the relation 
$X_2+\bar X_2=X_3+\bar X_3=0$, as was seen in (\ref{gauge1}). Therefore the target space of the dual sigma model is the 1-complex
dimensional space given by
\begin{eqnarray}
X_a&\in& \mathbb{C},\,  \,  \,   \sum_a n_a^2=1. \label{VacuumG1}\\
\sum_a X_a n_a&=&\frac{t}{2 Q},\,  \,   \,     \,      \,   X_2+\bar X_2=X_3+\bar X_3=0, \nn\\
\text{with quantum symmetry: }X_a n_a&\rightarrow&X_a n_a+\frac{2\pi i
 k_a}{2 Q},\, \, k_a\in \mathbb{Z}.\nn
\end{eqnarray}
This determines a family of dual manifolds given by the numbers $\{ n_a \}$. \footnote{\label{dualspace} Making the change of variables
$Y_a=e^{X_a n_a}$ one obtains the complex one dimensional space
$Y_a\in \mathbb{C},\,  \,Y_1 Y_2 Y_3=e^{\frac{t}{2 Q}},\,  \,   \,     \,      \,   |Y_2 |= |Y_3 |=1$. Quantum symmetry: $Y_a \sim Y_a e^{\frac{2\pi ik_a}{2 Q}},\, k_a\in \mathbb{Z}$.}

If we fix the gauge symmetry differently to obtain $\Phi_1=i\Phi_2$ and this will imply the relation 
$X_1+\bar X_1=X_3+\bar X_3=0$. Therefore the target space of the dual sigma model is the 1-complex
dimensional space given by
\begin{eqnarray}
X_a&\in& \mathbb{C},\,  \,  \,   \sum_a n_a^2=1. \label{VacuumG2} \\
X_a n_a&=&\frac{t}{2 Q},\,  \,   \,     \,      \,   X_1+\bar X_1=X_3+\bar X_3=0, \nn\\
\text{with quantum symmetry: }X_a n_a&\rightarrow&X_a n_a+\frac{2\pi i
 k_a}{2 Q},\, k_a\in \mathbb{Z}.\nn
\end{eqnarray}
 Both families of dual spaces (\ref{VacuumG1}) and (\ref{VacuumG2}) are equivalent under the permutation $1\leftrightarrow 2$. Next we are going to analyze the inclusion of a truly  non-Abelian part of the duality, which will modify
 the dual theory and it will affect the supersymmetric vacuum. However there is a non-Abelian subset of the $SU(2)$ duality group
 that gives rise to vacua coinciding with the ones of this Abelian family. This will be done by relaxing the condition that the expansion
 coefficients of the chiral superfields $n_a$ multiplying the $SU(2)$ generators are constant.

\section{Dualization in $SU(2)$ direction $n_a \sigma_a$, with semichiral $n_a$}
\label{Sec:SU2NA}

In this Section we go beyond a family of Abelian T-dualities inside the non-Abelian T-duality,
i.e.,  we consider a direction inside an $SU(2)$ gauged group with some restrictions on the
non-Abelian vector superfield. We implement these conditions by writing the vector superfield
as $V=|V|(x^\mu,\theta^\alpha,\bar\theta^{\dot \alpha}) \sigma_a \hat n_a(x^\mu,\theta^\alpha,\bar\theta^{\dot \alpha})$ with the field
$\hat n_a(x^\mu,\theta^\alpha,\bar\theta^{\dot \alpha})$ depending on the superspace coordinates.
The vector superfield  of the gauged group will have the more general form $V= \sigma_a V_a(x^\mu,\theta^\alpha,\bar\theta^{\dot \alpha})$. We impose the condition $D_- \hat n_a=0$ in order to obtain the dual theory in terms of twisted
chiral superfields; we take this condition for convenience but it would be interesting to study the possible choices more systematically.

When $D_{-}\hat n_a=0$  is fulfilled one can obtain for the Lagrange multiplier terms 
\begin{eqnarray}
\int d^4\theta\tr (\Psi \Sigma)=\int d^4\theta \frac{1}{2}X_a \hat n_a |V|.\nn
\end{eqnarray}
The field $\hat n_a$ is
a vector field and in addition satisfies $D_-\hat n_a=0$. This is a semichiral condition in addition to the reality condition.
In this Section we will first search for a vector superfield satisfying this semichiral condition.
Under this choice of $\hat{n}_a$,  we can further evaluate the Lagrange multiplier term in the action (\ref{nablsu2})
\begin{eqnarray}
\frac{1}{2}\int d^4\theta \tr ( \Psi \bar D_+ (e^{-V}D_{-} e^V))&=&\int d^4\theta \Psi_a \bar D_+(D_{-} |V| \hat n_a)\\
&=&-\int d^4\theta  \bar D_+\Psi_a(D_{-} |V| \hat n_a)\nn \\
&=&\int d^4\theta  \bar D_+D_-\Psi_a(|V| \hat n_a)=\int d^4\theta  X_a(|V| \hat n_a). \nn
 \end{eqnarray}
 With these ingredients, the dual action is composed of the action valid for a constant $\hat n_a$ (\ref{Ldual1}) with the addition of an extra term:
\begin{eqnarray}
L_{dual}&=&\int d^4\theta \sqrt{( X_1  + \bar X_1)^2+ (X_2+\bar X_2)^2+ (X_3 +  \bar X_3)^2}+\label{exT}\\
&+& \int d^4\theta  X_a \hat n_a \ln(\mathcal{K}(X_i,\bar X_i,n_j))+c.c.- \int d^4\theta\frac{1}{2 e^2}\bar{\Sigma}_0\Sigma_0\nn\\
&+&\frac{2 Q}{2}\int d\bar \theta^-d\theta^+\left(X_a n_a-\frac{t}{2 Q}\right) \Sigma_0+\frac{2 Q}{2}\int d\bar \theta^+d\theta^-\left(\bar X_a n_a-\frac{\bar t}{2 Q}\right) \bar \Sigma_0,\nn\\
&+&\frac{2 Q}{4}\int d\bar \theta^-d\theta^+\left(X_a \bar D_+n_a\right) D_-V_0+c.c.\nn
\end{eqnarray}
The last line in (\ref{exT}) is a truly non-Abelian part of the duality. It gives an interaction term between the
$U(1)$ gauge field and the dual twisted superfields. To evaluate this contribution, we need to compute also $D_-V_0$, which we will do in the WZ gauge. We will explore the extra terms that contribute to the effective scalar potential.

Let us discuss the symmetries of this action. It is invariant under $X_a\rightarrow X_a+\frac{2\pi i
 k_a}{2 n_a Q},\, k_a\in \mathbb{Z}$,  taking into account that the $\theta$ angle has periodicity i.e. $t\rightarrow t+2 \pi i$.
The first two lines of (\ref{exT}) are invariant as the real part of $X_a$ is invariant and $\mathcal{K}$ is real. The last two are invariant,
because they arise both from $-2 Q\int d\theta^4 X_a n_a V_0\rightarrow -\frac{2\pi i k_a}{2}\int d\theta^4 V_0=\frac{2\pi i k_a}{2}
\int d\bar \theta^-d\theta^+ \Sigma_0$. This corresponds to a displacement $t\rightarrow t-2 \pi i k_a$.
It is possible then to make an Ansatz for the instanton contributions in the dual theory as $e^{-X_a n_a}$ that
respects the symmetries. However, the effective twisted superpotential for $\Sigma_0$ of the dual theory fails to match with the one of the original theory. Therefore the mentioned instanton contribution, which leads to correct results in the family
of Abelian dualities, is not the correct expression for the non-Abelian case.
This question will be explored in a future work. 

 Let us work first with $\hat n_a$ satisfying $D_-\hat n_a=0$. Also, we choose the space-time coordinates $y=x+i \theta \sigma\bar\theta$ and $\bar y=x-i \theta \sigma\bar\theta$, \footnote{In 4D the variable change reads
$y^\mu=x^{\mu}+i\theta^{\beta} \sigma^{\mu}_{\beta\dot\beta}\bar \theta^{\dot \beta}$
and $\bar y^\mu=x^{\mu}-i\theta^{\beta} \sigma^{\mu}_{\beta\dot\beta}\bar \theta^{\dot \beta}$.} which
in the 2D theory are given by
\begin{eqnarray}
y^{0,3}&=&x^{0,3}-i(\pm\theta^+\bar\theta^++\theta^-\bar\theta^-),\\
\bar y^{0,3}&=&x^{0,3}+i(\pm\theta^+\bar\theta^++\theta^-\bar\theta^-),
\end{eqnarray}
For reference let us write the covariant derivatives of $y^{\mu}$ and $\theta$ coordinates, those are:
\begin{eqnarray}
D_{\alpha}\bar\theta^{\dot\beta}&=&\bar D_{\dot\alpha}\theta^{\beta}=0,\bar D_{\dot \alpha}\bar\theta^{\dot\beta}=-\delta^{\dot\beta}_{\dot \alpha}, D_{\alpha}\theta^{\beta}=\delta^{\beta}_{\alpha}, \\
D_{\alpha}\bar y^\mu&=&\bar D_{\dot\alpha}y^\mu=0,\nn\\
\bar D_+ \bar y^0&=&2 i \theta^+,\, \, \bar D_+ \bar y^3=-2 i\theta^+,\nn\\
\bar D_- \bar y^0&=&2 i \theta^-,\, \, \bar D_- \bar y^3=2 i\theta^-,\nn\\
D_+ y^0&=&-2 i \bar\theta^+,\, \, D_+  y^3=2 i \bar \theta^+,\nn\\
D_-  y^0&=&-2 i\bar\theta^- ,\, \, D_- y^3= -2 i \bar\theta^-.\nn
\end{eqnarray}
Let us write a generic vector superfield ($w$) expansion in terms of coordinates $y^{\mu}$. From the hermiticity condition $w=w^\dagger$ we have
\cite{wessbagger}
\begin{eqnarray}
w&=&C(y)+i \theta\gamma(y)-i \bar\theta\bar\gamma(y)+\frac{i}{2}\theta^2(M(y)+i N(y))-\frac{i}{2}\bar\theta^2(M(y)-i N(y))\label{wGauge}\\
&+&\theta \sigma^{\mu}\bar\theta w_{\mu}(y)+\bar\theta^2\theta\lambda(y)+\theta^2\bar \theta\bar\lambda(y)+\theta^2\bar\theta^2 D(y).\nn
\end{eqnarray}
In addition we require $D_- w=0$. The following Ansatz for a vector field $w=\hat n_a$ fullfils $D_{-}\hat n_a=0$ 
\begin{eqnarray}
\hat n_a&=&n_a(\bar y)-\theta^+\bar\theta^+  (w^a_0(\bar y)-w^a_3(\bar y))+i\theta^+ \gamma^a_+(\bar y)+i\bar\theta^+ \bar \gamma^a_+(\bar y).\label{exp}
\end{eqnarray}
One needs  to impose also $\hat n_a(\bar y)^{\dagger}=\hat n_a(\bar y)$. Performing the expansion of (\ref{exp}) one gets
\begin{eqnarray}
n_a(\bar y)&=&n_a(x)+i \theta^+\bar\theta^+(\partial_0-\partial_3)n_a+i\theta^-\bar\theta^-(\partial_0+\partial_3)n_a+\\
&+&2 \theta^+\bar\theta^+\theta^-\bar\theta^-(\partial_0^2-\partial_3^2)n_a,\nn\\
-\theta^+\bar\theta^+ w_{03}(\bar y)&=&\theta^+\bar\theta^+ w_{03}(x)-i\theta^+\bar\theta^+\theta^-\bar\theta^-(\partial_0+\partial_3)w_{03}(x),\nn\\
i\theta^+	\gamma^a_+(\bar y)&=&i\theta^+\gamma^a_+(x)-\theta^+\theta^-\bar\theta^-(\partial_0+\partial_3)\chi_+^a(x),\nn\\
i\bar\theta^+\bar\gamma^a_+(\bar y)&=&i\bar\theta^+\bar\gamma^a_+(x)-\bar\theta^+\theta^-\bar\theta^-(\partial_0+\partial_3)\bar\gamma_+^a(x).\nn
\end{eqnarray}
Recalling the reality condition the total $\hat n_a(\bar y)$ has the expression
\begin{eqnarray}
\hat n_a(\bar y)&=&n_a(x)
+2 \theta^+\bar\theta^+\theta^-\bar\theta^-(\partial_0^2-\partial_3^2)n_a(x)+
\theta^+\bar\theta^+ w_{03}^a(x)\nn\\
&+&i\theta^+\gamma^a_+(x)-\theta^+\theta^-\bar\theta^-(\partial_0+\partial_3)\gamma_+^a(x),\nn\\
&+&i\bar\theta^+\bar\gamma^a_+(x)-\bar\theta^+\theta^-\bar\theta^-(\partial_0+\partial_3)\bar\gamma_+^a(x).\nn
\end{eqnarray}
Let us compute now $\bar D_+ \hat n_a$ for subsequent use
\begin{eqnarray}
\bar D_+ \hat n_a(\bar y)&=&2i \theta^+ (\partial_0-\partial_3)n_a-\theta^+\theta^-\bar\theta^-(\partial_0^2-\partial_3^2)n_a+2\theta^+\theta^-\bar\theta^-(\partial_3^2-\partial_0^2)n_a+\label{D+n}\\
&-&\theta^+w^a_{03}(x)-i\theta^+\theta^-\bar\theta^- (\partial_0+\partial_3)w^a_{03}(x)-i \bar\gamma^a_+(x)\nn\\
&-&\theta^+\bar\theta^+(\partial_0-\partial_3)\bar\gamma^a_+(x)+\theta^-\bar\theta^-(\partial_0+\partial_3)\bar\gamma^a_+(x)
-i\theta^+\bar\theta^+\theta^-\bar\theta^-(\partial_0^2-\partial_3^2)\bar\gamma^a_+(x).\nn
\end{eqnarray}
Notice that even for $\hat n_a$ non constant, i.e., a non-Abelian direction in the vector superfield but $w^a_{03}=0$ the terms contributing to the scalar potential (not to the kinetic term) coincide with the ones for $\hat n_a$ constant discussed in Section \ref{Subsec:Ta}. The expression (\ref{D+n}) is simplified by recalling the reality condition of $\hat n_a(\bar y)$:
\begin{eqnarray}
\bar D_+ \hat n_a(\bar y)&=&2\theta^+\theta^-\bar\theta^-(\partial_3^2-\partial_0^2)n_a-\theta^+w^a_{03}(x)-i \bar\gamma^a_+(x)\nn\\
&-&\theta^+\bar\theta^+(\partial_0-\partial_3)\bar\gamma^a_+(x)
-i\theta^+\bar\theta^+\theta^-\bar\theta^-(\partial_0^2-\partial_3^2)\bar\gamma^a_+(x).\nn
\end{eqnarray}
Let us now write the vector superfield of the GLSM $U(1)$ gauge group $V_0$ in the Wess-Zumino gauge \cite{wessbagger}
\begin{eqnarray}
V_{0}&=&-\theta \sigma^{\mu}\bar\theta v_{\mu}(x)-i\bar\theta^2\theta\lambda(x)+i\theta^2\bar \theta\bar\lambda(x)+	\frac{1}{2}\theta^2\bar\theta^2 D(x),\\
&=& (v_{0}-v_3)\theta^+\bar \theta^++(v_3+v_0)\theta^-\bar\theta^--\sqrt{2}\bar\sigma_0\theta^-\bar \theta^+-\sqrt{2}\sigma_0\theta^+\bar \theta^-
\nn\\
&+&2 i\theta^+\theta^-(\bar\theta^-\bar\lambda^{+}-\bar\theta^+\bar\lambda^{-})+ 2 i\bar\theta^+\bar\theta^-(\theta^+\lambda^{-}-\theta^-\lambda^{+}).\nn\\
&+&2\theta^+\theta^-\bar\theta^+\bar\theta^- D(x).\nn
\end{eqnarray}
Furthermore the quantity $D_-V_0$ appearing in (\ref{exT}) and the field strength read:
\begin{eqnarray}
D_-V_0&=&\bar\theta^-(v_3+v_0)-i\bar\theta^-\theta^+\bar\theta^+(\partial_0+\partial_3)v_{30}+i\sqrt{2}\bar\theta^-\theta^-\bar\theta^+ (\partial_0+\partial_3)\bar\sigma_0\\
&-&\sqrt{2}\bar\sigma_0\bar\theta^+-2 i \theta^+\bar\theta^- \bar \lambda^++2 i\theta^+\bar \theta^+\bar\lambda^--2 i \bar\theta^+\bar\theta^-\lambda^+,\nn\\
&&+2 \theta^+\theta^-\bar\theta^+\bar\theta^-(\partial_0+\partial_3)\bar\lambda^--2\theta^+\bar\theta^+\bar\theta^- D,\nn \\
\Sigma_0&=&\frac{1}{2}\bar D_+ D_- V_0,\\
&=& \frac{1}{\sqrt{2}}\bar\sigma_0+i \theta^+\bar\lambda^-+i\bar\theta^-\lambda^++\theta^+\bar\theta^-(D-i (\partial_0 v_3-\partial_3v_0))\nn\\
&+&-i\frac{1}{\sqrt{2}}\bar\theta^-\theta^-(\partial_0+\partial_3)\bar\sigma_0-\frac{1}{\sqrt{2}}\theta^+\theta^-\bar\theta^+\bar\theta^-(\partial_0^2-\partial_3^2)\bar\sigma_0+2\theta^+\bar\theta^+\bar\theta^-(\partial_0-\partial_3)\lambda^+\nn \\
&-&2\theta^+\theta^-\bar\theta^-(\partial_0+\partial_3)\bar\lambda^-.\nn
\end{eqnarray}
In order to see if there are extra contributions to the scalar potential we calculate 
\begin{eqnarray}
-\frac{2 Q}{4} \int d\theta^+ d\theta^- d\bar\theta^- d\bar\theta^+ X_aV_0\hat n_a&=& - Q \sqrt{2}\bar\sigma_0 (2G_a n_a+ i\sqrt{2}\gamma_+^a\chi_-^a)+\\
&-&\frac{Q}{2}(v_3+v_0)(w_{03}^ax_a+i\sqrt{2}\bar\gamma_+^a\bar\chi_+^a)+Q D n_a x_a\nn\\
&-&Q \sqrt{2}i \bar\lambda^-\chi_-^a- Q i\sqrt{2}\lambda^+\bar\chi_+^a.\nn
\end{eqnarray}
Comparing with (\ref{abinside}), we see there is an extra contribution to the scalar potential, only dependent on the gauged field boson
$w_{03}^a$, the sum of the $U(1)$  vector potentials $v_3+v_0$ and the twisted chiral scalar $x_a$. This extra contribution is given by  $U=\frac{Q}{2} w_{03}^a(v_3+v_0)(x_a+\bar x_a)$.  This matches the observation that 
\begin{eqnarray}
\frac{2Q}{4}\int d\bar\theta^-d\theta^+  X_a \bar D_+ \hat n_a D_- V_0=-\frac{Q}{2} w_{03}^a x_a (v_0+v_3)+....\label{NAcontrib}
\end{eqnarray}

This renders a total scalar potential, in which we have included interaction terms with
the $U(1)$ gauge field component $v_0+v_3$. Via a gauge transformation this field combination can be set to a constant\footnote{Additionally note that
by going to the Higgs branch, with $\sigma_0=0$, due to the mass hierarchies in the IR \cite{Hori:2000kt} one can integrate $v_{03}$ and the field strength components do not contribute to the action.}, therefore we can look at the
effective potential obtained after taking the Higgs branch and integrating $v_{03}$:
\begin{eqnarray}
U=2 Q^2 e^2 |x_a n_a-t/(2Q)|^2+\frac{Q}{2} w_{03}^a (x_a+\bar x_a) (v_0+v_3).\label{Poteff2}
\end{eqnarray} 
The previous vacua found  from (\ref{Udual}) remains the same by taking $\forall_a w_{03}^a=0$. This is a restriction on the semichiral
real superfield $\hat n_a$ to $\hat n_a=n_a(\bar y)+i \theta^+\gamma_+(\bar y)+i	\bar\theta^+\bar\gamma_+(\bar y)$, excluding
total derivatives, which restricts the gauged field $V$.  A particular case is when $\hat n_a$
depends on space-time $\hat n_a=n_a(x)$. This is very interesting, because it implies that the results
of previous Section for a family of Abelian T-dualities inside of $SU(2)$ apply beyond the Abelian situation.
Because $\hat n_a$ is not constant one has really a vector superfield with the direction in the Lie algebra 
$V=|V|(x,\theta,\bar\theta)\hat n_a(x,\theta,\bar\theta)\sigma_a$.
Even if $\hat n_a$ has the discussed restrictions the components of $V$ along the different generators
$V_a=|V|(x,\theta,\bar\theta)\hat n_a(x,\theta,\bar\theta)$ are different from each other.  This lets us identify the vacua discussed in
Subsection \ref{Subsec:Ta} as vacua of a non-Abelian dual theory.

Furthermore we can consider  $w_{03}^a$ and $(v_0+v_3)$ to be different from zero, and investigate how the Higgs branch vacuum with
$\sigma_0=0$ is modified. Let us see this in the gauge $X_2+\bar X_2=X_3+\bar X_3=0$.  The
vacuum of (\ref{Poteff2}) is given by
\begin{eqnarray}
\Re(x_1)&=&-\frac{4 B_1 }{8 A n_1^2}- \frac{t_1}{2Q n_1}, \label{plane}\\
 \Im (x_3)&=& -\frac{\Im(x_1) n_1}{n_3} - \frac{\Im(x_2) n_2}{n_3} + \frac{t_2}{2Q n_3}, \nn
\end{eqnarray}
with $B_a=\frac{Q}{2}(v_0+v_3)\omega_{03}^a$ and $A=2 Q^2 e^2 $ such that the potential (\ref{Poteff2}) reads $U=A|x_a n_a-t/(2Q)|^2+B_a (x_a+\bar x_a)$.

Without fixing the gauge one obtains for the vacuum equations
\begin{eqnarray}
\Re(x_1)&=&-\frac{B_1 }{2 A n_1^2}-\frac{\Re(x_2)n_2}{n_1}-\frac{\Re(x_3)n_3}{n_1}- \frac{t_1}{2Q n_1},\\
 \Im (x_3)&=& -\frac{\Im(x_1) n_1}{n_3} - \frac{\Im(x_2) n_2}{n_3} + \frac{t_2}{2Q n_3}, \nn\\
 B_1 n_1&=&B_3 n_3.\nn
\end{eqnarray}
The addition of the non-Abelian contribution (\ref{NAcontrib}) changes the vacuum by changing the hyperplane to be
$x_a n_a=\frac{t}{2Q}-\frac{B_1 }{2 A n_1}$. The condition $\omega_{03}^1 n_1=\omega_{03}^3 n_3$
needs to be satisfied. In this vacuum there are two positive eigenvalues and four zero eigenvalues of the
Hessian matrix $(0,0,0,0,A,A)$.
The positive eigenvalues represent the growing directions of the potential. The zero eigenvalues of
the Hessian represents the real dimension 4 of the vacuum space.  
\begin{table}[htp]
\begin{center}
\begin{tabular}{|c|c|} \hline
$\lambda$& $Eigenvectors$ \\ \hline
0& $(0, 0, 0, -\frac{n_3}{n_1}, 0, 1)$ \\ \hline
0& $(0, 0, 0, -\frac{n_2}{n_1}, 1, 0)$ \\ \hline
0& $(-\frac{n_3}{n_1}, 0, 1, 0, 0, 0)$ \\ \hline
0& $(-\frac{n_2}{n_1}, 1, 0, 0, 0, 0)$ \\ \hline
A& $(0, 0, 0, \frac{n_1}{n_3}, \frac{n_2}{n_3}, 1)$ \\ \hline
A& $(\frac{n_1}{n_3}, \frac{n_2}{n_3}, 1, 0, 0, 0)$\\ \hline
\end{tabular}
\end{center}
\caption{Eigenvectors and eigenvalues of the Hessian matrix in the critical points of the Higgs branch.
The four zero eigenvalues corresponds to the real dimensions of the vacuum space,
directions in which the potential is flat.}
\label{eigenDual}
\end{table}

The directions of the eigenvector $(\frac{n_1}{n_3},\frac{n_2}{n_3},1,0,0,0)$ corresponds to non zero eigenvalue $A$. The infinitesimal transformation in the directions in which the potential grows are given by
\begin{eqnarray}
\Re(x_1) n_1&\rightarrow& \Re(x_1) n_1+ \delta z \frac{n_1}{n_3}	 n_1,\\
\Re(x_2) n_2&\rightarrow& \Re(x_2) n_2+ \delta z	\frac{n_2}{n_3} n_2,\nn\\
\Re(x_3) n_3&\rightarrow& \Re(x_3) n_3+ \delta z  n_3,\nn\\
\sum_a \Re(x_a) n_a&\rightarrow& \sum_a \Re(x_a) n_a+\frac{1}{n_3}\delta z.\nn
\end{eqnarray}

One direction corresponding to $0$ eigenvalue, the first in Table \ref{eigenDual}, gives a flat direction. This transformation preserves the condition (\ref{plane}) and is given by
\begin{eqnarray}
\Im(x_1)n_1&\rightarrow& \Im(x_1)n_1+(-\frac{n_3}{n_1} \delta y)n_1,\\
\Im(x_3)n_3&\rightarrow& \Im(x_3)n_3+(\delta y) n_3,\nn\\
\sum_a \Im(x_a) n_a&\rightarrow& \sum_a \Im(x_a) n_a. \nn
\end{eqnarray}

Fixing the gauge as in (\ref{gauge1}) the new vacuum is given by the 2 dimensional space:
\begin{eqnarray}
\sum_a x_a n_a&=&\frac{t}{2Q}-\frac{B_1 }{2 A n_1},\,  \, \label{geom2}\\
x_2+\bar x_2&=&x_3+\bar x_3=0.\nn
\end{eqnarray}
At the quantum level, i.e., the level of the partition function there is in addition the symmetry $x_a\rightarrow x_a+\frac{2\pi i
 k_a}{2 n_a Q},\, k_a\in \mathbb{Z}$ coming from the periodicity of $t$. The classical vacua are given by
 a plane in $\mathbb{R}^3$. Taking into account the quantum symmetry the supersymmetric vacua space
is given by a $T^2$ torus. The value of the potential in the minimum is given by
\begin{eqnarray}
U_{min}&=&(B_1 (-2 B_1 + A n_1 \Re(t_1)))/(8 A n_1^2).
\end{eqnarray}
To obtain a  supersymmetric minimum is required a zero scalar potential $U=0$, this can be achieved by setting $B_1=\frac{A n_1 \Re(t_1)}{2}$,
i.e. $(v_0+v_3)\omega_{03}^1=2 Q e^2 n_1 \Re(t_1)$.\footnote{\label{dualspace2}
As in footnote (\ref{dualspace}) one can make a change of variables trying to grasp better the geometry of the
one complex dimensional curve. The change of variables is given by $y_a=e^{x_a n_a}$, this gives
the one complex dimensional space
$y_a\in \mathbb{C}, \, \ y_1y_2y_3=e^{\frac{t}{8}-\frac{B_1 }{2 A n_1}},\,  \, |y_2|=|y_3|^2=1$.
The quantum shift symmetry becomes $y_a\sim y_a e^{\frac{\pi i k_a}{2 Q}},\, k_a\in \mathbb{Z}$.}

The geometry of the susy vacua space of the non-Abelian dual theory  (\ref{geom2}) and of the family of Abelian dual theories in (\ref{VacuumG1}) coincides. This is another indication that the family of Abelian dualities captures
features of the non-Abelian T-Duality. We also found in this Section that the family of Abelian dualities inside of $SU(2)$ describes really
a subset of the full non-Abelian duality. We obtained a Lagrangian where the coefficient $\hat n_a$
determining the vector superfield can variate for example the space-time, giving for the vector superfield components the dependence $V_a=|V|(x^{\mu},\theta^{\alpha},\bar\theta^{\dot\alpha})n_a(x^{\mu})$. This means that the scalar potential coincides in both cases.  
Nevertheless there are kinetic terms which make the Lagrangians distinct. When we look at the vacuum we considered the  $\hat n_a$ as
constant coefficients, but they appear in the action and make it invariant under local transformations
that are not Abelian. Therefore further investigation should me made to clarify the connection. If one considers a more general expansion, then an additional interaction term
between $U(1)$ vector superfields of the GLSM and twisted chiral scalars arise (\ref{NAcontrib}), that deform this vacuum. The general conclusion is that in this model a family of Abelian dualities inside the gauged group can capture
relevant features of the non-Abelian T-duality. The two dimensional space obtained as supersymmetric vacuum
has the same definition in both cases, with only a change in the effective $t$ parameter. However
the true comparison should be made after including the instanton corrections. We need to find the
expression for the instanton contributions in the non-Abelian dual theory.
It would be interesting to check if for other groups a family of Abelian dualities also casts important properties of 
the non-Abelian duality.

\section{Comparison with mirror symmetry and Abelian T-duality}
\label{Sec:Analysis}

In this Section we analyze different dualities of the $\mathbb{CP}^1$ model obtained from
a $U(1)$ GLSM with $SU(2)$ global symmetry, i.e., two chiral superfields with equal charges, which has as target
variety $\mathbb{CP}^1$. We first compare the Abelian dualization of the $\mathbb{CP}^1$ model leading to
mirror symmetry as done by Hori and Vafa with our dualization procedure. Then we compare a family of Abelian dual models which
 is inside the non-Abelian T-duality  with a  model obtained by a single $U(1)$ dualization.  We have already shown that  the family of Abelian dual models inside $SU(2)$ are more than just  Abelian, displaying properties of the full non-Abelian duality.  The mirror theory
obtained with two Abelian dualities and the non-Abelian dualization lead to different  geometries.  However,
a more careful comparison will have to include the knowledge of the instanton contributions
in the non-Abelian dual theory, we discuss also the steps to investigate this question. 




The initial theory is the  $\mathbb{CP}^1$ model given by a $U(1)$ GLSM with two equally charged chiral superfields $\Phi_1$, $\Phi_2$
described in Subsection \ref{SU2glsm}, and we want to describe its mirror model. The Hori-Vafa dualization is performed on the two
chiral superfields $\Phi_1$, $\Phi_2$ to obtain the twisted chiral superfields
$Y_1$ and $Y_2$. These two
Abelian dualities lead to a dual model with a twisted superpotential 
$W(\Theta) \sim \Theta+1/\Theta$ \cite{Hori:2000kt}. This superpotential gives rise to Toda $A_1$ space $\{z\in\mathbb{C}^*,W(z):=z+1/z\}$.
It is appropriate to integrate the $U(1)$ gauge field strength in the IR limit  since the mass
of $\sigma_0$, the scalar component of $\Sigma_0$, is high due to the
relation of the coupling and the energy scale $e\gg \Lambda$. The dual model obtained by Hori-Vafa is given by the Lagrangian
\begin{eqnarray}
L_{mirror}&=& -\frac{1}{2}\int d\theta^4(Y_1+\bar Y_1)\ln (Y_1+\bar Y_1)-\frac{1}{2}\int d\theta^4(Y_2+\bar Y_2)\ln (Y_2+\bar Y_2)\\
&+&\frac{1}{2}\int d\tilde \theta^2 \Sigma_0(Y_1+Y_2-t)+\frac{1}{2}\int d\tilde \theta^2 (e^{-Y1}+e^{-Y_2})+h.c.\nn\\
&-&\frac{1}{2 e^2}\int d \theta^4 \bar \Sigma_0 \Sigma_0.\nn
\end{eqnarray}

First, we consider the vacua obtained in the Higgs branch without taking into account the instanton corrections. Taking into account the K\"ahler metric components $K_{1\bar 1}=-\frac{1}{2(Y_{1}+\bar Y_{1})}$, $K_{2\bar 2}=-\frac{1}{2(Y_{2}+\bar Y_{2})}$, and 
the auxiliary fields contribution $K_{i\bar j}G_i\bar G_{\bar j}$ from the kinetic term the Lagrangian leads to the scalar potential
\begin{eqnarray}
U &=&-(G_1 + G_2) \sigma_0 - (\bar G_1 + 
\bar G_2) \bar \sigma_0 - \frac{1}{2 (y_1 + \bar y_1)} G_1 \bar G_1 - \frac{1}{2 (y_2 + 
\bar y_2)} G_2 \bar G_2\nn \\
&-& (y_1 + y_2 - t) (D - i v_{03}) - (\bar y_1 + \bar y_2 - \bar t) (\bar D + i \bar v_{03}) - \frac{1}{2 e^2} |D - i v_{03}|^2. 
\end{eqnarray}
Let us integrate out the auxiliary fields $G_1$, $G_2$ and the field combination $D-i v_{03}$ to obtain:
\begin{eqnarray}
U_{eff}= 2 e^2 |y_1+y_2-t|^2+|\sigma_0|^2 2 (y_1+y_2+\bar y_1+\bar y_2).\label{UeffToda}
\end{eqnarray}

The solution is given by $y_1+y_2-t=0$ and $\sigma_0=0$. This is equivalent to considering the classical solution and go to the Higgs branch, integrating $\Sigma_0$,  which in the IR limit develops a large mass. In the case when we only consider the potential for the dynamical fields in the GLSM, integrating out $G_1,G_2$
and the real auxiliary field $D$ one obtains the potential
\begin{eqnarray}
U = 2 |\sigma_0|^2(y_1 + \bar y_1 + y_2+\bar y_2) + 
 \frac{e^2}{2} ( y_1+\bar y_1+ y_2 + \bar y_2-t - \bar t )^2.
\end{eqnarray}
This leads to a Higgs branch at $\sigma_0=0$ and $\Re(y_1+y_2)=\Re(t)=r$. On the other hand the Coulomb branch appears at $\Re(y_1+y_2)=0, \sigma_0=2 e \sqrt{r}$, which is supersymmetric only at $r=0$, therefore there is no supersymmetric Coulomb branch.

The interaction terms of $v_{03}$ with the scalars are added in order to determine the effective potential:
\begin{eqnarray}
U_{eff,0}&=&-\frac{1}{2 e^2}(v_{03}^2 + 2 i e^2 v_{03} (t - \bar t - y_1 + \bar y_1 - y_2 + \bar y_2)\label{intTod} \\
&-&2 e^4 (-t - \bar t + y_1 + \bar y_1 + y_2+ \bar y_2)^2).\nn
\end{eqnarray}
In the Higgs branch the $\Sigma$ field gets a large mass, and integrating it out is equivalent to integrating out $v_{03}$ from (\ref{intTod}) to obtain (\ref{UeffToda}).
Therefore the condition $Y_1+Y_2=t$ comes from integrating out $\Sigma_0$  along the  Higgs branch of the theory. The relevant twisted superpotential of the effective theory is given by $\widetilde W=e^{-Y_1}+e^{-t+Y_1}$, or after making the change of variables
$\Theta=t/2-Y_1$, one obtains $\widetilde W=e^{-t/2}(e^{-\Theta}+e^{\Theta})$. This is a Landau-Ginzburg model with
twisted superpotential $\widetilde W$, constituting the $A_1$ Toda supersymmetric field theory. It has two vacua with scalar component of $\Theta$ given by $\phi_{\Theta}=\pi i=2 \pi i$, 
these can be mapped to the $\mathbb{CP}^1$ model.

Let us compare the mirror theory obtained with the Hori-Vafa procedure to the one obtained by our dualization procedure. This corresponds to the case discussed in Subsection  \ref{multipleSF}. We specialize to the case of three chiral superfields, two of them equally charged.
After dualization there will be a total Lagrangian given by:
\begin{eqnarray}
L_{2}&=&L_{mirror}+\int d\theta^4 \bar\Phi_{3} e^{2 Q_{0,3} V_0}\Phi_{3}.
\end{eqnarray}
This is the specialization of Lagrangian (\ref{LdualM}) to the mentioned case. There is an spectator chiral superfield $\Phi_3$ which is uncharged under the gauged symmetry. The scalar potential of the dual theory is given by
\begin{eqnarray}
U_2&=& -\left[(y_1+\bar{y}_1) +(y_2+\bar{y}_2) -(t+\bar{t})\right] D - 2 \Re[ \sigma_0 (G_1+G_2)] - \frac{1}{2e^2} D^2\label{mir2}\\
&+&K_{i \bar{j}} G_{i} \bar{G}_{\bar{j}}-F_3 \bar{F_3} - Q_{0,3}D x_3\bar{x}_3,\nn
\end{eqnarray}
with $F_3$ and $x_3$ the auxiliary and scalar component of $\Phi_3$ respectively. The scalar potential (\ref{mir2}) has a Higgs branch with vevs $\sigma_0=x_3=0$, where the $U(1)$ field can be
integrated out. This leads to the effective potential $U_{eff}=2 e^2 |t - y_1 - y_2|^2$.  The resulting effective
theory is obtained by integrating $\Sigma_0$ and is given by the condition $Y_1+Y_2=t$ 
and twisted superpotential $\widetilde W=e^{-Y_1}+e^{-t+Y_1}$. This is exactly the mirror symmetric theory
obtained as well from the Hori-Vafa's procedure discussed before.

The mirror theory of the $\mathbb{CP}^1$ model is different from the dual theories considered in our work. In particular it is
different from the family of Abelian dualities inside of $SU(2)$, because to obtain it the dualization is done along two $U(1)$ directions. Therefore, it is natural to expect a difference when one compares with the non-Abelian duality.  However, a true comparison with the non-Abelian case requires the non-Abelian instanton corrections. From the non-Abelian dual theory we know the supersymmetric vacua given in (\ref{geom2}), but instanton corrections will change this picture. Hori and Vafa studied the Euclidean Wick rotated original GLSM finding the
instanton solutions in their Subsection 3.2. To match with the dual theory they study a fermionic two-point function $\langle  \chi_+ \bar\chi_-\rangle$ with the fermion components
of the dual field $Y$ in the original and the dual theory.
The result is matched by identifying the $e^{-Y}$ contribution to the twisted superpotential in the dual theory. To determine
the non-Abelian instanton correction to $\widetilde W$ we would need to study similar objects in the non-Abelian dual theory.
The Ansatz $e^{-X_a n_a}$ obtained for a family of Abelian dualities is an starting point, but
as it was discussed in Section \ref{Sec:SU2NA} is not the correct modification since it does not lead to the correct effective
$U(1)$ superpotential on the non-Abelian dual theory.

Next, let us match the family of Abelian dualities inside $SU(2)$ with an $U(1)$ dualization of the theory. Considering the instanton Ansatz $e^{-X_a n_a}$ the family is given by a constant twisted superpotential. This should fit with the model obtained by
a single $U(1)$ dualization. We compare the dual model obtained dualizing the $U(1)$ GLSM with two equal charged chiral superfields (\ref{lab}) along 
an $U(1)$ direction of one of them, with the model of Abelian duality inside $SU(2)$ presented in Section \ref{Subsec:Ta} ($\ref{nabl}$).
The kinetic part of ($\ref{nabl}$) can be written in matrix form as $\bar\Phi^T M \Phi$, where $M=e^{|V| n_a\sigma_a}$ and $\Phi^T = \Phi $, then it is possible to rotate the element $M$ into an element generated by $\sigma_1$ to obtain
\begin{eqnarray}
e^{V_1  \sigma_1 }= \left( \begin{array}{cc}
    e^{V_1} + e^{-V_1} & e^{V_1}- e^{-V_1}\\ 
e^{V_1}- e^{-V_1} & e^{V_1}+ e^{-V_1} \\ 
  \end{array}\right) .
\end{eqnarray}  
This can be seen from formula (\ref{AbsV1}). Then the kinetic term in ($\ref{lab}$) can be written as
\begin{equation}
\frac{1}{2}\left( \bar\Phi_1 \bar\Phi_2\right) \left( \begin{array}{cc}
    e^{V_1} + e^{-V_1} & e^{V_1}- e^{-V_1}\\ 
e^{V_1}- e^{-V_1} & e^{V_1}+ e^{-V_1} \\ 
  \end{array}\right)\left( \begin{array}{c}
  \Phi_1\\ 
\Phi_2 \\ 
  \end{array}\right)  =\bar\Phi_1' e^{2 Q_0 V_0+V_1} \Phi_1' + \bar\Phi_2' e^{2 Q_0 V_0 -V_1} \Phi_2',\label{kinS1}
\end{equation}
where the prime fields are obtained by diagonalizing the exponential of the vector superfield matrix to get $\Phi_1'=(\Phi_1+\Phi_2)/\sqrt{2}$ and $\Phi_2'=(\Phi_2-\Phi_1)/\sqrt{2}$. Since the GLSM has an $U(1)$ symmetry, we can make a gauge transformation $V_0 \rightarrow V_0 + \frac{V_1}{2 Q_0}$ which is
substituted in (\ref{kinS1}) to obtain
\begin{equation}
\bar\Phi_1' e^{2 Q_0 V_0+2V_1} \Phi_1' + \bar\Phi_2'e^{2 Q_0 V_0} \Phi_2'. \;  
\end{equation}
Setting the charge $Q$ to $1$ this is exactly the kinetic term of the two chiral superfields GLSM ($\ref{lab}$).  Also the Lagrange multiplier term is given by $\int d^4 \theta \tr( \Lambda \Sigma)=\int d^4 \theta X_1 V_1$. These observations establish that the gauged theories are equivalent. Now, let us compare the vacuum of both models. In the model of duality along the direction $\sigma_1$ in $SU(2)$ the vacuum is given at $x_1=	\frac{t}{2Q}$ with constant twisted superpotential 
$\widetilde W=2\mu e^{-t/(2Q)}$. On the dualization of the Abelian symmetry along one chiral superfield the susy vacuum is given at $y=t/Q$ and the twisted superpotential reads $\widetilde W=e^{-t/Q}$. These results are equivalent after taking into account the correct normalization factors.

The family of Abelian dualities inside of $SU(2)$ leads to a dual model with scalar potential that coincides with the one of
the models with more general vector superfields (\ref{exp}). These duals with more generic vector superfields constitute
truly non-Abelian duals. The extension to more general vector superfields gives rise to extra terms in the Lagrangian (\ref{exT}).
In this non-Abelian case one has to still determine the  instanton correction to $\widetilde W$. This will be a next step in order to explore the non-Abelian T-duality in GLSMs leading to determinantal CY varieties and their connection to mirror symmetry.  

In this Section we first compared the mirror of the $\mathbb{CP}^1$ GLSM obtained with the Hori-Vafa procedure,
to the one obtained by our dualization procedure. We also checked that non-Abelian T-duality coincides in the limit with Abelian T-duality,
by comparing the resulting Lagrangians and the effective theories. This is an important sanity check and a necessary step toward testing the duality. With the methods developed we plan to consider in the future the study of non-Abelian T-dualities in more general examples.

\section{Conclusions}
\label{Sec:Conclusions}
We have used the method of gauging a global symmetry in field theories, with subsequent addition of a Lagrange multiplier enforcing flatness condition and alternative elimination of fields, to  describe T-duality in GLSMs. Tackling the description of T-Duality in GLSMs by this gauging procedure allows us to obtain a Lagrangian with added Lagrange multipliers
and a vector superfield. Integration of the latter leads to the dual model. The method applies to GLSMs with multiple U(1)'s and multiple chiral superfields. More importantly for our goals is that the method admits a  generalization to the gauging of non-Abelian global symmetries, which lead to non-Abelian-T dualities in GLSMs.

Let us emphasize that  with respect to the Hori-Vafa dualization procedure our method has the benefit of being easily generalized for the case of non-Abelian symmetries. Recall that the Hori-Vafa's method consists on creating a Lagrangian
that upon integration of certain fields leads to one or the dual model. In the case when the global symmetry is non-Abelian there is not known systematic way of constructing the analog of Hori-Vafa master Lagrangian. We have, however, presented a systematic way of generating such master Lagrangian using the method studied in this manuscript. Our method can also be applied to GLSMs with non-Abelian
gauge groups, as the ones leading to determinantal CY varieties.  To the best of our knowledge, our results constitute the first systematic formulation of non-Abelian T-duality in $(2,2)$ gauged linear sigma models. 

Considering a $U(1)$ GLSM with $n$ chiral superfields and a generic non-Abelian global symmetry group we gauged the global symmetry.  We obtained the equations of motion for this generic case. The non-Abelian duality leads to twisted superfields only for certain cases, constituting an important difference with the Abelian T-duality.  For generic groups the elementary fields in the equations of motion are semi-chiral.

 We have analyzed in detail the concrete example of a GLSM with a single $U(1)$ gauge field and two chiral superfields of equal charges.  This theory has a global $SU(2)$ symmetry which can be gauged. For  $SU(2)$ group the fundamental fields of the duality are twisted chiral superfields. The K\"ahler potential of the dual theory is obtained in terms of the dual twisted
 superfields.  First, assuming an Abelian direction inside of the gauged group the  twisted superpotential can be obtained as well.
 This direction means that we consider a vector superfield proportional to a linear combination
 of $SU(2)$ generators with constant coefficients. This constitutes a family of Abelian dualities which
display characteristics of the full non-Abelian duality, as is supported by the obtained K\"ahler potential.  Furthermore, 
for coefficients depending only on a  scalar component the scalar potential reduces
to the one of a family of Abelian dualities. This observation implies 
that the afore mentioned family of Abelian dualities is richer than one could have naturally expected since it describes
a truly non-Abelian duality inside of $SU(2)$, even if not the most general one.  In a next step
we generalize the dualization to truly non-Abelian T-Duality by imposing a semi-chiralilty condition on the coefficients of the generator expansion. 
The supersymmetric vacuum of this non-Abelian dual model  has the same geometry as the supersymmetric vacuum of the family of Abelian T-dualities previously mentioned. Both vacua definitions differ only by a constant coming from an interaction
term between the gauged fields and the GLSM original $U(1)$ gauge field. In the IR limit, {\it i.e.},  considering the Higgs branch when the $U(1)$ field strength is integrated out, this term contributes to the scalar potential, and therefore to the supersymmetric vacuum. 

We took some steps toward the inclusion of nonperturbative effects. Namely, we postulated an Ansatz for the twisted-superpotential instanton corrections.  In the dual theory along an Abelian direction of the group (which is a linear combination of the generators) we verified that the effective action for the $U(1)$ gauge field coincides for the  original and the dual model. These instanton corrections allow to compare the Abelian dual theory family inside the non-Abelian dual theory
with the case of an Abelian T-dual model with a single $U(1)$ dualization. The result is that they match, {\it  i.e.},  their effective potentials have the same vacuum
and the twisted superpotential can be mapped to each other. For the fully non-Abelian case however, we do not have the instanton corrections; this is an important direction that we leave for future investigations.

It is worth pointing out a number of interesting problems that our investigation naturally highlights. With the methods studied here it would be interesting to tackle some explicit examples of dualities for determinantal CY varieties such as those presented in \cite{Rødland_2000}. In particular, it seems possible to arrive at a more systematic description of mirror symmetry in the Pfaffian CYs  presented in \cite{Kanazawa:2012xya}, which were recently realized as GLSMs \cite{Caldararu:2017usq}. Note that explicit  computations of Gromov-Witten invariants for the CY of \cite{Rødland_2000}  were given in \cite{Hosono:2007vf}, providing a concrete framework for testing the non-Abelian T-duality proposed in this manuscript. Given the advances in explicit computation in GLSMs, it would be interesting to apply the technique of supersymmetric localization to our models in order to better understand the effects of non-Abelian T-duality. 

Finally, we have seen glimpses of new representations, such as the semi-chiral superfield representation appearing in our construction. It would be useful to achieve a complete understanding of the possible representations, arguably  opening a window into a formulation of non-Abelian T-duality for generalized geometry. 

\section{Acknowledgments}
We thank Alejandro Cabo Bizet, Hugo Garc\'{\i}a Compe\'an, Vishnu Jejjala, Yulier Jim\'enez Santana, Albrecht Klemm, Oscar Loaiza Brito, Octavio Obreg\'on, Fernando Quevedo, Yongbin Ruan and Catherine Whiting for useful discussions and comments. NGCB thanks LCTP for warm hospitality; also she would like to thank the String Theory Group and MITP of Wits U.  where part of this research was developed for an exciting scientific environment; she acknowledges the support of the NRSF of South Africa, Project PRODEP UGTO PTC-515 and Project  CIIC 154/2018 {\it Teor\'{\i}as efectivas provenientes de la teor\'{\i}a de cuerdas: fenomenolog\'{\i}a y aspectos formales}.  LAPZ is partially supported by DoE  Grant No. de-sc0017808 {\it Topics in the AdS/CFT Correspondence: Precision tests with Wilson loops, quantum black holes and dualities}. The work of RSS is partially supported by CONACyT retention program, CONACyT Grant No. 271904 and Project PRODEP UDG-PTC-1368. 


\newpage
\appendix

\newpage

\section{Equations of GLSM non-Abelian T-dualization}
\label{appA}
Here we present details of the calculations in Subsection \ref{Sec:NATDeom} where we compute the equations of motion of the 
integrated vector superfield of the gauged symmetry.

In the following we expand the variations of the Lagrange multiplier terms in the Lagrangian. The variation with respect to the gauged field $V$ of the Lagrangian term for the Lagrange multiplier $\Psi$ is given by:
\begin{eqnarray}
\delta \tr (\bar D_{+} \Psi e^{-V} D_{-} e^V)&=&\tr\left((\chi \bar D_{+} \Psi+\bar D_{+} \Psi\chi +D_-\bar D_{+} \Psi)\Delta V\right),\\
&=&\tr((\chi \bar D_{+} \Psi+\bar D_{+} \Psi\chi +D_-\bar D_{+} \Psi)T_a)\Delta V_a,\nn\\
&=&\tr((\chi T_b T_a-T_b\chi T_a +T_b T_aD_-)\bar D_{+} \Psi_b\Delta V_a,\nn\\
&=&( \tr( e^{-V} D_{-} e^V [T_b,T_a])\bar{D}_{+} \Psi_b+D_{-}\bar{D}_{+}\Psi_a/2 )\Delta V_a \nn \\
&=&(i f_{abc}\tr( e^{-V} D_{-} e^V T_c)\bar{D}_{+} \Psi_b+D_{-}\bar{D}_{+}\Psi_a/2 )\Delta V_a. \nn
\end{eqnarray}
The hermitian conjugate term reads:
\begin{eqnarray}
\delta \tr (D_{+} \bar\Psi e^{V} \bar D_{-} e^{-V})&=&\tr\left(( -\bar D_{-} e^{-V} D_{+} \bar\Psi e^V+e^{-V} D_{+}\bar \Psi\bar D_{-} e^V +e^{-V}(D_+\bar D_{-} \Psi)e^V)\Delta V)\right),\nn\\
&=&\tr( -\bar D_{-} e^{-V} D_{+} \bar\Psi e^V+e^{-V} D_{+}\bar \Psi\bar D_{-} e^V +e^{-V}(D_+\bar D_{-} \Psi)e^VT_a)\Delta V_a,\nn\\
&=&( \tr(-e^V T_a\bar D_{-} e^{-V}T_b - D_{-} e^V T_ae^{-V} T_b)D_{+}\bar \Psi_b\nn \\
&+&\tr(e^{-V}T_be^VT_a)(D_+\bar D_{-} \Psi_b)\Delta V_a\nn\\
&=&( \bar D_{-}\tr(-e^V T_ae^{-V}T_b)\bar D_{+}\bar\Psi_b+\tr(e^{-V}T_be^VT_a)(D_+\bar D_{-} \Psi_b)\Delta V_a,\nn\\
&=&(D_+\bar D_{-} \Psi_b+D_+\bar \Psi_b \times \bar D_{-})\tr(e^V T_ae^{-V}T_b)\Delta V_a.
\end{eqnarray}
These expressions are valid for any group, and we write them for further reference.

\section{Formulae for $SU(2)$ group}
\label{SU2formulas}

In this Appendix we present relevant formulae for superfields calculations in the case of  the $SU(2)$ group. We start
presenting relations among the generators, formulae for the exponentials, and
expansions for the vector superfields. We also integrate the equations of motion for the vector superfield (\ref{teq3}).

For the generators $T_a$ of $SU(n)$ gauge the following identity applies: 
\begin{equation}
T_aT_b=\frac{1}{2n}\delta_{ab} 1+\frac{1}{2}(i f_{abc}+d_{abc})T_c.\label{TaTb}
\end{equation}

For the case $SU(2)$ lets us give the specification of (\ref{TaTb}) and another useful formula: 
\begin{eqnarray}
\sigma_a\sigma_b&=&\delta_{ab} 1+i \epsilon_{abc}\sigma_c,\\
\tr(\sigma_a\sigma_b\sigma_c)&=&2 i \epsilon_{abc}.\nn
\end{eqnarray}
The real vector superfield can be written as
\begin{eqnarray}
V=V_a\sigma_a=|V|\hat n\cdot \hat{\sigma}.\label{VaApp}
\end{eqnarray}
where $\sigma_a$ are the Pauli matrices, $\hat\sigma=(\sigma_1,\sigma_2,\sigma_3)$,  $\hat n_a=V_a/|V|$ is an unitary vector and $|V|=\sqrt{V_a V_a}$.

Performing the series expansion of the exponential one gets
\begin{eqnarray}
e^V&=&\cosh|V|+\hat n_a \sigma_a \sinh |V|,\\
e^{-V}&=&\cosh|V|-\hat n_a \sigma_a \sinh |V|,\nn\\
D_{-} e^{\pm V}&=&\left((\sinh|V|\pm\hat n_a \sigma_a \cosh |V|)D_{-}|V|
\pm D_{-}\hat{n}_a\sigma_a \sinh|V|\right),\nn\\
\bar D_{-} e^{\pm V}&=&\left((\sinh|V|\pm\hat n_a \sigma_a \cosh |V|)\bar D_{-}|V|
\pm \bar{D}_{-}\hat{n}_a\sigma_a \sinh|V|\right).\nn
\end{eqnarray}
Previous equations combine to give
\begin{eqnarray}
e^{-V}D_{-} e^V&=&\hat n\cdot \hat\sigma D_{-} |V|+D_{-}\hat n\cdot \hat{\sigma} \sinh|V| \cosh|V|-\hat n_{a} D_{-} \hat n_b \sigma_a \sigma_b \sinh |V|^2,\\
&=& (\hat n_a D_{-} |V|+D_{-} \hat n_a \sinh |V| \cosh|V|-\hat n_d D_{-} \hat{n}_b i \epsilon_{d b a} \sinh|V|^2)\sigma_a,\nn\\
&=&\chi_a \sigma_a. \nn
\end{eqnarray}
For the case $V=V_3 \sigma_3$  ones has $e^{-V}D_{-} e^V= D_- V_3 \sigma_3$ with $\tr e^{-V}D_{-} e^V=0$,
but $\tr  \bar D_+\Psi(e^{-V}D_{-} e^V)=\bar D_+\Psi_3 D_- V_3$.

There is a particular case easy to treat, this is when $V=V_a \hat n_a \sigma_a$ with $\hat n_a$ independent
of the superspace coordinates or satisfiying the restriction $D_{-} \hat{n}_a=0$. Then let us consider $V= |V|\hat n_a \sigma_a$ giving $e^{-V}D_{-} e^V=D_{-} |V| \hat n_a \sigma_a$. For this
case $\tr \Psi \Sigma=\bar D_+\Psi_a D_- |V| \hat n_a$. The explicit form of the exponential $e^V$
in this case is
\begin{eqnarray}
e^V&=&
\left(
\begin{array}{cc}
 \cosh|V|+n_3   \sinh|V|&(n_1+i n_2) \sinh |V|  \\
 (n_1-i n_2) \sinh |V|&    \cosh|V|-n_3   \sinh|V|
\end{array}
\right).
\end{eqnarray}
One can then use the relations (\ref{VvsPhi}) to obtain
\begin{eqnarray}
&&\Phi_1 \bar \Phi_2 \cosh |V| + ((n_1+ i n_2) |\Phi_1|^2 - 
 n_3 \Phi_1 \bar \Phi_2) \sinh  |V| + \label{AbsV1}\\
 &&\frac{X_1+\bar X_1-i (X_2+\bar X_2)}{2 e^{2 Q V_0}}=0.\nn
\end{eqnarray}
and
\begin{eqnarray}
\bar \Phi_1\Phi_2 \cosh |V| + ((n_1- i n_2) |\Phi_2|^2 +
 n_3 \bar \Phi_1 \Phi_2) \sinh  |V| \label{AbsV2}\\
 + \frac{X_1+\bar X_1+i (X_2+\bar X_2)}{2 e^{2 Q V_0}}=0.\nn
\end{eqnarray}
Setting $n_1=n_2=0$ the only possible solution implies $n_3=0$ also. So for the gauge
$V_1=V_2=0$ also $V_3=0$ is required. On the other hand $n_3=0$ implies
that $|\Phi_1|^2=|\Phi_2|^2$ meaning $\frac{\Phi_1}{\Phi_2}=\frac{\bar\Phi_2}{\bar \Phi_1}$. 
Let us study first the case $n_3=0$, in this case by summing up the equations (\ref{AbsV1}) and (\ref{AbsV2}) one gets:
\begin{eqnarray}
&&(\bar \Phi_1\Phi_2 + \Phi_1\bar\Phi_2)\cosh |V| +n_3(\bar \Phi_1\Phi_2 - \Phi_1\bar\Phi_2)\sinh |V| \label{AbsV3}\\
&+& 2n_1|\Phi_1|^2  \sinh  |V| + \frac{X_1+\bar X_1}{e^{2 Q V_0}}=0.\nn
\end{eqnarray}
and by subtracting them:
\begin{eqnarray}
&&(\bar \Phi_1\Phi_2 - \Phi_1\bar\Phi_2)\cosh |V|+n_3(\bar \Phi_1\Phi_2 + \Phi_1\bar\Phi_2)\sinh |V|\label{AbsV32} \\
&-&i 2 n_2 |\Phi_1|^2\sinh  |V|   + i \frac{X_2+\bar X_2}{e^{2 Q V_0}}=0.\nn
\end{eqnarray}
For convenience let us rewrite (\ref{P1P2}) as
\begin{eqnarray}
\bar \Phi_1\Phi_2=\frac{\Phi_2}{\Phi_1}|\Phi_1|^2&=&|\Phi_1|^2\frac{-(X_3+\bar X_3)\pm \sqrt{(X_3+\bar X_3)^2+((X_1+\bar X_1)^2+(X_2+\bar X_2)^2)})}{(X_1+\bar X_1- i (X_2+\bar X_2)) }\nn \\
&=&|\Phi_1|^2 F(X_a,\bar X_a).
\end{eqnarray}
The expression for $F$ is given by
\begin{eqnarray}
F(X_a,\bar X_a)&=&-\frac{\left(X_3+\bar X_3\pm \sqrt{\sum_a(X_a+\bar X_a)^2}\right)(X_1+\bar X_1+i(X_2+\bar X_2))}{(X_1+\bar X_1)^2+(X_2+\bar X_2)^2},\label{Fexpr}\\
\Re F=\frac{F+\bar F}{2}&=&-\frac{\left(X_3+\bar X_3\pm \sqrt{\sum_a(X_a+\bar X_a)^2}\right)(X_1+\bar X_1)}{(X_1+\bar X_1)^2+(X_2+\bar X_2)^2},\nn \\
\Im F=\frac{F-\bar F}{2 i}&=&-\frac{\left(X_3+\bar X_3\pm \sqrt{\sum_a(X_a+\bar X_a)^2}\right)(X_2+\bar X_2)}{(X_1+\bar X_1)^2+(X_2+\bar X_2)^2}. \nn
\end{eqnarray}

Thus obtaining from (\ref{AbsV3})
\begin{eqnarray}
(F+\bar F)\cosh |V| + (2n_3(F-\bar F)+2n_1) \sinh  |V| + \frac{X_1+\bar X_1}{e^{2 Q V_0}|\Phi_1|^2 }=0.\label{AbsV4}
\end{eqnarray}
This is solved in terms of $|V|$ to obtain
\begin{eqnarray}
|V|&=&\ln\left(\frac{-\frac{X_1+\bar X_1}{e^{2 Q V_0}|\Phi_1|^2 }+\sqrt{4 n_1^2-(F+\bar F)^2+\left(\frac{X_1+\bar X_1}{e^{2 Q V_0}|\Phi_1|^2 }\right)^2}}{F+\bar F+ 2 n_1}\right),\label{AbsVSol1}\\
&=&\ln\left(\frac{-(X_1+\bar X_1)+\sqrt{(4 n_1^2-(F+\bar F)^2)(e^{2 Q V_0}|\Phi_1|^2)^2+\left(X_1+\bar X_1\right)^2}}{F+\bar F+ 2 n_1}\right)+\\
&-&2 Q V_0-\ln(|\Phi_1|^2 ).\nn
\end{eqnarray}
As well from (\ref{AbsV32}) one gets
\begin{eqnarray}
i 2\Im F\cosh |V|+(n_3(F+\bar F) -i 2 n_2) \sinh  |V|  + i \frac{X_2+\bar X_2}{e^{2 Q V_0}|\Phi_1|^2}=0.\label{AbsV33}
\end{eqnarray}
This is solved  by
\begin{eqnarray}
|V|&=&\ln\left(\frac{-\frac{X_2+\bar X_2}{e^{2 Q V_0}|\Phi_1|^2 }+\sqrt{4 n_2^2-4 \Im F^2+\left(\frac{X_2+\bar X_2}{e^{2 Q V_0}|\Phi_1|^2 }\right)^2}}{2 \Im F-2 n_2}\right),\label{AbsVSol2}\\
&=&\ln\left(\frac{-(X_2+\bar X_2) +\sqrt{(4 n_2^2-4 \Im F^2)(e^{2 Q V_0}|\Phi_1|^2)^2+\left(X_2+\bar X_2\right)^2}}{2 \Im F-2 n_2}\right)+\\
&-&2 Q V_0-\ln(|\Phi_1|^2 ).\nn
\end{eqnarray}
One can equate (\ref{AbsVSol1}) and (\ref{AbsVSol2}) to obtain an expression for $(e^{2 Q V_0}|\Phi_1|^2)$. This will serve to finally eliminate
the chiral and antichiral superfields $\Phi_1$ and $\bar \Phi_1$ from the action. The result reads
\begin{eqnarray}
&&\sinh |V| ((n_2+i n_3/2(F+\bar F))(X_1+\bar X_1)+(n_1+n_3/2(F-\bar F))(X_2+\bar X_2))\\
&-&\cosh |V| ((X_1+\bar X_1)\Im F-(X_2+\bar X_2)\Re F)=0.\nn
\end{eqnarray}
This relation implies either $n_1=n_2=0$, $n_1=X_1=0$, $n_2=X_2=0$ or $X_1=X_2=0$. The
most interesting possibilities seem to be the 2nd and the 3rd. The solution with $V_2=V_3=X_2=0$,
$V=V_1$ and $|\Phi_1|^2=|\Phi_2|^2$ implies $2(X_1+\bar X_1)(X_3+\bar X_3)=0$ which is an
additional restriction on $X_1$ and $X_3$. As well $V_1=V_3=X_1=0$,
$V=V_2$ and $|\Phi_1|^2=|\Phi_2|^2$ implies $2(X_2+\bar X_2)(X_3+\bar X_3)=0$ which is an
additional restriction on $X_2$ and $X_3$.

A solution for $|V|$ can also be obtained for generic values of $n_1,n_2$ and $n_3$ 
using (\ref{AbsV1}) and (\ref{AbsV2}). This is given by
\begin{eqnarray}
|V|&=&\ln(\mathcal{K}(X_i,\bar X_i,n_j))-2 Q V_0-\ln (|\Phi_1|^2),\nn
\end{eqnarray}
The expression $\mathcal{K}(X_i,\bar X_i,n_j)$ in the argument of the logarithm is given by:
{\footnotesize
\begin{eqnarray}
\mathcal{K}&=&\frac{(\bar F - n_1 - i n_2 + \bar F n_3) (X_1  + \bar X_1+ i (X_2 + \bar X_2)) }{2 F((\bar F -1) n_1 - i (\bar F^2+1) n_2 + 2 \bar F n_3)}\label{KLog}\\
&+&\frac{F ( \bar F (n_1 - i n_2) + n_3-1) (X_1+\bar X_1 - i (X_2+ \bar X_2)))}{2 F((\bar F -1) n_1 - i (\bar F^2+1) n_2 + 2 \bar F n_3)}.\nn
\end{eqnarray}
}

\section{Twisted Chiral Expansion}
\label{twistedExpansion}
In this Section we compute explicitly some useful expression in components  of the (anti-) twisted chiral superfields used along the paper, that
can become handy for working with twisted chiral superfields. These derivations
are based on the procedure  for chiral superfields presented in the book of Wess and Bagger \cite{wessbagger}. 

First, let us recall the expansions of twisted and anti-twisted chiral superfields  
 \begin{eqnarray}
Y_i&=&y_i+\sqrt{2}\theta^+ \bar \chi_+ +\sqrt{2}\bar \theta^{-}\chi_-+2 \theta^+\bar \theta ^- G_i+...,\label{tc}\\
\bar Y_i&=&\bar y_i+\sqrt{2}\chi_+\bar\theta^+  +\sqrt{2}\bar \chi_-\theta^{-}+2   \theta ^-\bar\theta^+  \bar G_i+... \label{atc}
\end{eqnarray}
In previous formulae $y_i$, $\chi_-,\bar \chi_+$, $G_i$ represent the scalar, fermionic and auxiliary components of the twisted
chiral superfield $Y_i$. Their conjugates are the components of the (anti-) chiral superfield $\bar Y_i$. Consider a function $K$ representing the K\"ahler potential that depends on twisted chiral and anti-twisted chiral fields $Y_i$ and $\bar Y_{i}$, where $i$  runs from $1,\ldots, n$. Now, assuming that $K(Y_1, \ldots Y_n, \bar Y_1, \ldots, \bar Y_n)$ can be Taylor expanded, we compute a generic term of the expansion namely,
\begin{equation}
\label{gen}
Y_{i_1} \cdots Y_{i_N} \bar Y_{j_1} \cdots \bar Y_{j_M}.
\end{equation}
Since this generic term can be regarded as the product of the monomial $Y_{i_1} \cdots Y_{i_N}$ and $\bar Y_{j_1} \cdots \bar Y_{j_M}$, thus let us Taylor expand the functions $P(Y_{i_1} \cdots Y_{i_N})$ and $\bar P (\bar Y_{j_1} \cdots \bar Y_{j_M})$ to obtain
\begin{eqnarray}
P(Y_{i_1} \cdots Y_{i_N})= P(y) + \sqrt{2} \theta^+ \bar \chi_{+ \, i} \frac{\partial P(y)}{\partial y_i} + \theta^{+} \bar \theta^{-} \left[ G_i \frac{\partial P(y)}{\partial y_i} -\frac{1}{2} \bar{\chi}_{+\, i} \chi_{- \, j} \frac{\partial^2 P(y)}{\partial y_i \partial y_j}\right] , \\
\bar P(\bar Y_{j_1} \cdots \bar Y_{j_M})= \bar P(\bar y) + \sqrt{2} \chi_{+\, i} \bar{\theta}^{+}  \frac{\partial \bar P(\bar y)}{\partial \bar{y}_i} + \theta^{-} \bar \theta^{+} \left[ \bar{G}_i \frac{\partial \bar{P}(\bar{y})}{\partial \bar{y}_i} -\frac{1}{2} \bar{\chi}_{+\, i} \chi_{- \, j} \frac{\partial^2 \bar{P}(\bar{y})}{\partial \bar{y}_i \partial \bar{y}_j}\right],\nn
\end{eqnarray}
where $y=(y_{i_1},\ldots, y_{i_N}) $.  We can use this expressions since $K$ can be regarded as products $Y$'s and $\bar Y$'s

\begin{eqnarray}
& &P(Y)P(\bar Y)= P \bar P + 2 \theta^{-} \chi_{+ \, j} \bar{\theta}^{+}  \frac{\partial}{\partial \bar y_j} P \bar{P} + \theta^{-} \bar{\theta}^{+} \left[ \bar{G}_i \frac{\partial}{\partial \bar y_j} P \bar{P} - \frac{1}{2} \chi_{+ \, i} \bar{\chi}_{- \, j} \frac{\partial^2}{\partial \bar{y}_i \partial \bar{y}_j} P \bar{P} \right] \nn \\ 
& &+ \sqrt{2} \theta^{+} \bar{\chi}_{+ i} \frac{\partial}{\partial y_i} P \bar{P} + 2 \theta^{+} \bar{\theta}^{+} \bar{\chi}_{+ \, i} \chi_{+ \, j} \frac{\partial^2}{\partial y_i\partial \bar y_j} P \bar{P}  \\
& & +\sqrt{2} \theta^{+} \theta^{-} \bar\theta^{+} \chi_{+ \, i} \left[\bar{G}_j \frac{\partial^2}{\partial y_i \partial \bar y_j}  - \frac{1}{2} \chi_{+ \, j} \bar\chi_{k} \frac{\partial^3}{\partial y_i \partial \bar y_j \partial \bar y_k} \right] P \bar{P} \nn\\
&& +\theta^{+} \bar\theta^{-} \left[ G_i \frac{\partial}{\partial \bar y_i}  -\frac{1}{2} \bar\chi_{+ \, i} \chi_{-\, j} \frac{\partial^2}{\partial \bar y_i \partial \bar y_j} \right]  P \bar{P} \nn \\ 
& &-\sqrt{2} \bar \theta^{+} \theta^{+} \bar \theta^{-} \chi_{+ k} \left[ G_i \frac{\partial^2}{\partial y_i\partial \bar y_k} - \frac{1}{2} \bar{\chi}_{+ \, i} \chi_{- \, j} \frac{\partial^3}{\partial y_i \partial y_j \partial \bar y_k}  \right] P \bar{P} \nn \\ & & +\theta^{+} \bar{\theta}_{-} \theta^{-} \bar \theta^{+}  \Big[ G_j \bar{G}_i \frac{\partial^2}{\partial \bar{y}_i \partial y_j}  -\frac{1}{2} G_i \chi_{+ \, j} \bar{\chi}_{- k} \frac{\partial^3}{\partial y_i \partial \bar y_j \partial \bar{y}_k} -\frac{1}{2} \bar{G}_k \bar\chi_{+\, j} \chi_{-\, j} \frac{\partial^3}{\partial y_i \partial \bar y_j \partial y_k} \nn \\& &+\frac{1}{4}  \bar\chi_{+\, i} \chi_{-\, j} \chi_{+\, k} \bar{\chi}_{-\, \ell} \frac{\partial^4}{\partial y_i \partial y_j \partial \bar{y}_k \partial \bar{y}_\ell} \Big] P \bar{P}\nn \label{pp}
\end{eqnarray}

The contribution to the Lagrangian is the term proportional to $\theta \theta \bar \theta \bar \theta$, which is easily computed in $(\ref{gen})$ with the aid of the previous expression setting $P = Y_{i_1} \cdots Y_{i_N}$ and $\bar P =  \bar Y_{j_1} \cdots \bar Y_{j_M}$, without including fermionic terms we get
\begin{eqnarray}
Y_{i_1} \cdots Y_{i_N} \bar Y_{j_1} \cdots \bar Y_{j_M}= \cdots +4 \theta^{+} \bar{\theta}^{-} \theta^{-} \bar{\theta}^{+} \left[ G_i \bar{G}_{\bar{i}} \frac{\partial}{\partial y_i} (y_{i_1} \cdots y_{i_N}) \frac{\partial}{\partial \bar{y}_i} (\bar{y}_{\bar{i}_1} \cdots \bar{y}_{\bar{i}_N}) \right] \nn \\
= \cdots + 4 \theta^{+} \bar{\theta}^{-} \theta^{-} \bar{\theta}^{+} \left[ G_i \bar{G}_{\bar{i}} \frac{\partial^2}{\partial y_i  \partial \bar{y}_i} (y_{i_1} \cdots y_{i_N} \bar{y}_{\bar{i}_1} \cdots \bar{y}_{\bar{i}_N}) \right], \label{genterm}
\end{eqnarray} 
hence the $\theta \theta \bar \theta \bar \theta$ term of $K(Y_{i_1}, \ldots, Y_{i_n}, \bar Y_{j_1}, \ldots ,\bar Y_{j_n})$ is the sum over all the monomials (with their respective constants of the power expansion) of the form $(\ref{genterm})$. Thus the K\"ahler potential contribution reads
\begin{eqnarray}
K(Y_{i_1}, \ldots, Y_{i_n}, \bar Y_{j_1}, \ldots ,\bar Y_{j_n})&=& \cdots + 4 \theta^{+} \bar{\theta}^{-} \theta^{-} \bar{\theta}^{+} \left[ G_i \bar{G}_{\bar{j}} \frac{\partial^2}{\partial y_i  \partial \bar{y}_{j}} K(y_1,\ldots, y_{i_1}, \bar{y}_{\bar{1}}, \ldots, \bar{y}_{\bar{i}_n}) \right] \nn \\ 
&=& \cdots + 4 \theta^{+} \bar{\theta}^{-} \theta^{-} \bar{\theta}^{+}  G_i \bar{G}_{\bar{j}} K_{i \bar{j}}  .
\end{eqnarray}
In general if $P= y_{i_1} \cdots y_{i_N}$ and $\bar P = \bar y_{j_1} \cdots \bar y_{j_M}$, summing over all the terms of  the expansion of the K\"ahler potential together with the following results of K\"ahler geometry
\begin{eqnarray}
K_{i \bar j}&=& \frac{\partial}{\partial y_i \partial_{\bar{j}}} K , \\
K_{i \bar j \, , k}&=& \frac{\partial}{\partial y_k}  K_{i \bar j} =K_{\ell \bar j} \Gamma^{\bar{\ell}}_{i k},\\
K_{i \bar j \, , \bar{k}}&=&\frac{\partial}{\partial \bar{y}_k}  K_{i \bar j} =K_{i \bar \ell} \Gamma^{\bar{\ell}}_{\bar{j} \bar{k}},
\end{eqnarray}
we can express (\ref{pp}) as:
\begin{eqnarray}
K(Y,\bar{Y}) &=& K - 2 \theta^{-}  \bar{\theta}^{+}  \chi_{+ \, j}  \frac{\partial}{\partial \bar y_j} K + \theta^{-} \bar{\theta}^{+} \left[ \bar{G}_i \frac{\partial}{\partial \bar y_j} K- \frac{1}{2} \chi_{+ \, i} \bar{\chi}_{- \, j}  K_{\bar{i} \bar{j}} \right]  \\ 
&+& \sqrt{2} \theta^{+} \bar{\chi}_{+ i} \frac{\partial}{\partial y_i} K + 2 \theta^{+} \bar{\theta}^{+} \bar{\chi}_{+ \, i} \chi_{+ \, j}  \frac{\partial^2}{\partial \bar y_i \partial \bar y_j}K  \nn 
\\ &+&\sqrt{2} \theta^{+} \theta^{-} \bar\theta^{+} \chi_{+ \, i} \left[\bar{G}_j K_{i \bar{j}} - \frac{1}{2} \chi_{+ \, j} \bar\chi_{k}  K_{i \ell} \Gamma^{\ell}_{\bar j \bar k} \right] \nn\\
&+& \theta^{+} \bar\theta^{-} \left[ G_i \frac{\partial}{\partial \bar y_i} K -\frac{1}{2} \bar\chi_{+ \, i} \chi_{-\, j} \frac{\partial^2}{\partial \bar y_i \partial \bar y_j} K \right] \nn \\ 
&-&\sqrt{2} \bar \theta^{+} \theta^{+} \bar \theta^{-} \chi_{+ \, k} \left[ G_i  K_{i \bar{k}} - \frac{1}{2} \bar{\chi}_{+ \, i} \chi_{- \, j}  K_{j \bar \ell} \Gamma^{\bar \ell}_{\bar k i} \right]   \nn
 \\ &+& \theta^{+} \bar{\theta}_{-} \theta^{-} \bar \theta^{+} \Big[ G_j \bar{G}_i K_{i \bar{j}} -\frac{1}{2} G_i \chi_{+ \, j} \bar{\chi}_{- k} K_{i \bar \ell} \Gamma^{\bar \ell}_{\bar j \bar k} -\frac{1}{2} \bar{G}_k \bar{\chi}_{+ \, j} \chi_{-\, j} K_{j \bar \ell} \Gamma^{\bar \ell}_{\bar{k} i} \nn \\ 
 &+&\frac{1}{4} \bar\chi_{+\, i} \chi_{-\, j} \chi_{+\, k} \bar{\chi}_{-\, \ell} K_{j \bar{k}, \, i \bar{\ell} } \Big]. \nn
\end{eqnarray}

\section{Dualization along $\sigma_3$, vector superfield: $|V|=V_3$}
In this Appendix  we discuss the dualization along a particular direction of the $SU(2)$ global symmetry,
this is the direction along the $\sigma_3$ generator. This case was not considered in the main text
because it has the particularity that solving the equations of motion for the vector superfield
in terms of the twisted chiral superfields one encounters the condition $V_3=0$. Nevertheless, 
one can write the dual model Lagrangian, and compare the suspersymmetric vacuum
with the one obtained in the case of Abelian duality.

Let us consider that the $e^V$ components fullfill $e^V_{12}=e^V_{21}=0$. This is to choose a direction along $\sigma_3$,
which constitutes a reduction to an Abelian model.
For this case
\begin{eqnarray}
e^{V}=e^{V_3 \sigma_3}= \left( \begin{array}{cc}
    e^{-V_3} & 0 \\ 
    0 & e^{V_3} \\ 
  \end{array}\right) ,\,  |V|=V_3,\, n_3=1.
\end{eqnarray}
The two following relations coming from the equation of motion for $V$ (\ref{teq3}) hold
\begin{eqnarray}
e^{2 Q V_0} \bar\Phi_1e^V_{11}\Phi_1&=&(X_1+\bar X_1- i (X_2+\bar X_2))\frac{\Phi_1}{2 \Phi_2 },\label{despeje10}\\
e^{2 Q V_0} \bar\Phi_2 e^V_{22}\Phi_2&=&(X_1+\bar X_1+ i (X_2+\bar X_2))\frac{\Phi_2}{2  \Phi_1 }.\nn
 \end{eqnarray} 
 These imply
\begin{eqnarray}
e^{2 Q V_0} e^V_{11}&=&(X_1+\bar X_1- i (X_2+\bar X_2))\frac{1}{2 \Phi_2 \bar \Phi_1},\label{despeje1}\\
e^{2 Q V_0} e^V_{22}&=&(X_1+\bar X_1+ i (X_2+\bar X_2))\frac{1}{2  \Phi_1\bar \Phi_2 }.\nn
 \end{eqnarray} 
which is a bit problematic because in this case $e^V_{11}=1/e^V_{22}$. 
 Let us look at the other term in the Lagrangian
 \begin{eqnarray}
 \tr(\Psi\Sigma)= -\frac{1}{2}\bar D_+\Psi_3 D_- V_3=-\frac{1}{2}(-D_-(\bar D_+\Psi_3 V_3)+X_3 V_3)=-\frac{1}{2}X_3 V_3,
  \end{eqnarray} 
from this term a twisted superpotential is generated.

On the other hand from (\ref{despeje1}) we obtain
\begin{eqnarray} 
 e^V_{11}=e^{-V_3}&=&\frac{(-(X_1+\bar X_1)- i (X_2+\bar X_2))}{2 \bar \Phi_1\Phi_2e^{2 Q V_0} },\\
V_3&=&-\ln \frac{(-(X_1+\bar X_1)- i (X_2+\bar X_2))}{2 \bar \Phi_1\Phi_2e^{2 Q V_0} },\nn\\
&=&-\ln\frac{-(X_1+\bar X_1)^2-(X_2+\bar X_2)^2}{-(X_3+\bar X_3)^2\pm \sqrt{\sum_a (X_a+\bar X_a)^2}}+\ln (2 \bar \Phi_1\Phi_1)+2 QV_0.\nn
\end{eqnarray}
There is an apparent problem, which is that using the component $e^V_{22}$ expression
in (\ref{VvsPhi}) one gets a result for $V_3$ which implies that $V_3=-V_3\rightarrow V_3=0$. 
\footnote{This observation can be connected with the fact that in the WZ gauge one has for the Lagrange multiplier term:
$\int d^4\theta \tr ((\chi \tau+\tau \chi+D_- \tau )\sigma_3)$$=\int d^4\theta \tr ((\chi \tau+\tau \chi+D_-\bar D_+ \Psi )\sigma_3)$$=\int d^4\theta \tr (( \chi \tau+\tau \chi)\sigma_3)=0$. As well writing  explicitly (\ref{sol1}) as
$e^{2 Q V_0}\bar \Phi_i e^V_{ij}\Phi_j=$$-e^{2 Q V_0}\frac{2 \bar\Phi_1 \Phi_2}{e^V_{21}} - \frac{e^V_{22} (D_-\bar D_+ \Psi_1+ i D_-\bar D_+ \Psi_2+c.c.)}{e^V_{21}}  - D_-\bar D_+ \Psi_3+c.c$, the last term will drop when integrating over $d^4 \theta$. } As
however $V_0\neq 0$ and $\ln (2 |\Phi_1|^2)\neq 0$ one can write the dual model. The dual model is given by the Lagrangian
\begin{eqnarray}
L_{dual}&=&\int d^4\theta \sqrt{( X_1  + \bar X_1)^2+ (X_2+\bar X_2)^2+ (X_3 +  \bar X_3)^2}+\\
&+&\frac{1}{2}  \int d^4\theta  X_3 \ln\frac{(X_3+\bar X_3)^2\pm \sqrt{\sum_a (X_a+\bar X_a)^2}}{(X_1+\bar X_1)^2+(X_2+\bar X_2)^2}+c.c.\nn\\
&+&2 Q\int d\bar \theta^-d\theta^+\left(X_3-\frac{t}{4}\right) \Sigma_0+2 Q\int d\bar \theta^+d\theta^-\left(\bar X_3-\frac{\bar t}{4}\right) \bar \Sigma_0- \int d^4\theta\frac{1}{2 e^2}\bar{\Sigma}_0\Sigma_0.\nn
\end{eqnarray}

Let us briefly discuss that the dualization along $\sigma_3$ can be mapped to an Abelian dualization along the phase of one chiral superfield
as the one discussed in Subsection \ref{Subsec:1csf} and in Section \ref{Sec:Analysis}. in the main text this discussion is done for the direction $\sigma_1$ which is the one analyzed systematically. The kinetic part of ($\ref{nabl}$) can be written in matrix form as $\bar\Phi^T M \Phi$, where $M=e^{|V| n_a\sigma_a}$ and $\Phi^T = \Phi $. It is possible to rotate the element $M$ into an element generated by $\sigma_3$ to obtain
\begin{eqnarray}
e^{|V|  \sigma_3 }= \left( \begin{array}{cc}
    e^{|V|} & 0 \\ 
   0 & e^{-|V| } \\ 
  \end{array}\right) .
\end{eqnarray}  
Then the kinetic term in ($\ref{nabl}$) can be written as
\begin{equation}
L_{kin}=\bar\Phi_1 e^{2 Q_0 V_0+|V|} \Phi_1 + \bar\Phi_2 e^{2 Q_0 V_0 - |V|} \Phi_2. \label{kinS3}
\end{equation}
Since the GLSM has an $U(1)$ symmetry, we make a gauge transformation $V_0 \rightarrow V_0 + \frac{|V|}{2 Q_0}$ which is
substituted in (\ref{kinS3}) to obtain
\begin{equation}
L_{kin}=\bar\Phi_1 e^{2 Q_0 V_0+2|V|} \Phi_1 + \bar\Phi_2 e^{2 Q_0 V_0} \Phi_2 \; , 
\end{equation}
setting the charge $Q$ to $1$ this is exactly the kinetic term of the two chiral superfields GLSM ($\ref{lab}$).

\section{Notation and identities}
\label{notation}

Here we give a short resume of many of the conventions for the supersymmetry calculations
performed in the paper.

The sigma matrices, constituting $SU(2)$ generators are given by:
\[
\sigma^0=\left(
\begin{array}{cc}
 -1 & 0  \\
 0 & -1       
\end{array}
\right), \sigma^1=\left(
\begin{array}{cc}
0 & 1  \\
 1 & 0       
\end{array}
\right),\\
\sigma^2=\left(
\begin{array}{cc}
 0 & -i  \\
 i & 0      
\end{array}
\right), \sigma^3=\left(
\begin{array}{cc}
1 & 0  \\
0 & -1      
\end{array}
\right).
\]

Spinor indices are raised and lowered via the Levi-Civita tensor \cite{wessbagger}:
\begin{eqnarray}
\psi^{\alpha}=\epsilon^{\alpha\beta}\psi_\beta,\,      \,      \, \psi_{\alpha}=\epsilon_{\alpha\beta}\psi^\beta.
\end{eqnarray}
Under this transformation the superspace coordinates are related via $\theta_+=\theta^-$ and $\theta_-=-\theta^+$.
The conjugation of spinor products is performed via \cite{wessbagger}:
\begin{eqnarray}
(\chi \psi  )^\dagger=(\chi^{\alpha}\psi_\alpha)^\dagger=\bar\psi_{\dot\alpha}\bar\chi^{\dot\alpha}=\bar\psi \bar\chi=\bar\chi \bar\psi .
\end{eqnarray}

The following identities are of use in the derivations presented in Section 5. First we write some relations needed to reduce a
component of the vector superfield i.e. the gauge field in (\ref{wGauge}):
\begin{eqnarray}
\theta \sigma^m \bar\theta v_m&=&(v_3(y)-v_0(y))\theta^+\bar\theta^++(v_1(y)-iv_2(y))\theta^-\bar\theta^+\\
&+&(v_1(y)+iv_2)\theta^+\bar\theta^--(v_3(y)+v_0(y))\theta^-\bar\theta^-,\nn \\\
(\theta \sigma^m\bar \theta v_m)^\dagger&=&(v_3(\bar y)-v_0(\bar y))\theta^+\bar\theta^++(v_1(\bar y)-iv_2(\bar y))\theta^-\bar\theta^+\\
&+&(v_1(\bar y)+iv_2(\bar y))\theta^+\bar\theta^--(v_3(\bar y)+v_0(\bar y))\theta^-\bar\theta^-,\nn\\
\theta\chi&=&\theta^\alpha\chi_\alpha=\theta^+\chi_++\theta^-\chi_-,  \\
\bar\theta\bar\chi&=&\bar\theta_{\dot\alpha}\bar\chi^{\dot\alpha}=-\bar\theta^+\bar\chi_+-\bar\theta^-\bar\chi_-.
\end{eqnarray}
Then we write the reduction of the fermionic components $\lambda,\bar \lambda$ of the vector superfield:
\begin{eqnarray}
i\theta^2\bar \theta\bar  \lambda(x)&=&i\theta^\alpha\theta_\alpha \bar\theta_{\dot \alpha}\bar\lambda^{\dot \alpha},\\
&=& 2 i\theta^+\theta^-(\bar\theta^-\bar\lambda^{+}-\bar\theta^+\bar\lambda^{-}),\nn \\
-i \bar \theta^2\theta \lambda(x)&=&-i \bar\theta_{\dot \alpha}\bar\theta^{\dot \alpha}\theta^\alpha\lambda_\alpha,\\
&=& 2 i\bar\theta^+\bar\theta^-(\theta^+\lambda^{-}-\theta^-\lambda^{+}).\nn
\end{eqnarray}

The integrations in superspace of the K\"ahler potential and the twisted superpotential are given by the conventions \cite{Hori:2000kt}:
\begin{eqnarray}
\int d^4 \theta K&=&\frac{1}{4}\int d \theta^+ d \theta^- d \bar \theta^- d \bar \theta^+ K,\nn \\
\frac{1}{2}\int d^2 \tilde \theta \widetilde W&=&\frac{1}{2}\int d \bar \theta^- d  \theta^+ \widetilde W,\nn \\
\frac{1}{2}\int d^2 \tilde \theta\overline{\widetilde W}&=&\frac{1}{2}\int d \bar \theta^+ d  \theta^- \overline{\widetilde W},\nn \\
\end{eqnarray}
In the notation from \cite{Witten:1993yc} one gets the following expression for the twisted chiral field strength:
\begin{eqnarray}
\Sigma&=&\frac{1}{\sqrt{2}}\bar D_+ D_- V =\sigma-i \sqrt{2}\theta^+\bar \lambda_+ -i\sqrt{2}\bar\theta^-\lambda_- + \sqrt{2}\theta^+\bar\theta^- (D-iv_{03})
\\
&-& i\bar\theta^-\theta^-(\partial_1-\partial_0)\sigma - i\theta^+\bar\theta^+(\partial_0 + \partial_1)\sigma + 
\sqrt{2}\bar\theta^-\theta^+\theta^-(\partial_0 -\partial_1)\bar\lambda_+  \nn\\
&+& \sqrt{2}\theta^+\bar\theta^-\bar\theta^+(\partial_0 +\partial_1)\lambda_- -\theta^+\bar\theta^-\theta^-\bar\theta^+(\partial_0^2 - \partial_1^2)\sigma.\nn
\end{eqnarray}
These formulas are useful for most of the superfield calculations performed in the main text.

\section{An $SU(2)$ scalar example, dualizing along $\sigma_3$ direction}
\label{appF}

The aim of this Appendix  is to work out  explicitly another example of gauging a global symmetry on a gauge theory.
We consider two complex scalar fields $\phi_1$ and $\phi_2$ charged with identical charge under an $U(1)$ gauge group with field $\hat A_{\mu}$ and the charge of both been $\hat q$.  This theory has an $SU(2)$
global symmetry.

Let us make a dualization of this symmetry only under the $\sigma_3$ generator direction. 
This is equivalent to having the two complex scalar fields of the doublet charged with opposite 
charges under another $U(1)$, and to dualize in that direction.  The corresponding covariant derivatives act on the fields as
\begin{eqnarray}
D_\mu \phi_1= ( \hat D_{\mu} + i q A_\mu)\phi_1,\ \  D_\mu \phi_2= ( \hat D_{\mu} - i q A_\mu)\phi_2.
\end{eqnarray}
Both fields have the same charge under the initial  $U(1)$ gauge so that $\hat D_\mu=\partial_\mu+i \hat q \hat A_\mu$.
The Lagrangian once the extra $U(1)$ symmetry has been gauged  is given by
\begin{eqnarray}
L_0=(D_\mu \phi_1)^\dagger D^\mu \phi_1+(D_\mu \phi_2)^\dagger D^\mu \phi_2+\lambda (\partial_0 A_1-\partial_1 A_0).
\label{sigma3L}
\end{eqnarray}
To obtain the dual model we integrate (\ref{sigma3L}) with respect to $A_\mu$ to obtain
\begin{eqnarray}
A^{\mu}=\frac{\partial_{\rho} \lambda \epsilon^{\rho \mu}-i q ((\hat D^\mu \phi_2) \phi_2^*-(\hat D^\mu \phi_2)^*\phi_2-(\hat D^\mu \phi_1) \phi_1^*+(\hat D^\mu \phi_1)^*\phi_1)}{2 q^2 |\phi|^2}.\label{AvsPhi}
\end{eqnarray}

Substituting the equation (\ref{AvsPhi}) in (\ref{sigma3L}) one obtains
\begin{eqnarray}
L=(\hat D_\mu \phi_1)^\dagger \hat D^\mu \phi_1+(\hat D_\mu \phi_2)^\dagger \hat D^\mu \phi_2-q^2 A_\mu A^\mu |\phi|^2.
\end{eqnarray}

The term $-q^2 A_0^2|\phi|^2$ will give
\begin{eqnarray}
-q^2 A_0^2|\phi|^2=-\frac{(\partial_1 \lambda)^2}{4 q^2|\phi|^2}+\frac{(\hat D^0 \phi_2\phi_2^*-\hat D^0 \phi_1\phi_1^*-h.c.)^2}{4 |\phi|^2}-\frac{i \partial_1 \lambda(\hat D^0 \phi_2\phi_2^*-\hat D^0 \phi_1 \phi_1^*-h.c.)}{2 q |\phi|^2}.\nn
\end{eqnarray}

The term $q^2 A_1^2|\phi|^2$ will give
\begin{eqnarray}
q^2 A_1^2|\phi|^2=\frac{(\partial_0 \lambda)^2}{4 q^2|\phi|^2}-\frac{(\hat D^1 \phi_2\phi_2^*-\hat D^1 \phi_1\phi_1^*-h.c.)^2}{4 |\phi|^2}-\frac{i \partial_0 \lambda(\hat D^1 \phi_2\phi_2^*-\hat D^1 \phi_1 \phi_1^*-h.c.)}{2 q |\phi|^2}.\nn
\end{eqnarray}
Such that:
\begin{eqnarray}
-q^2 A_\mu A^\mu |\phi|^2&=&\frac{\partial_\mu \lambda \partial^\mu \lambda}{4 q^2|\phi|^2}-\frac{i \partial^\mu\lambda  \epsilon_{\mu\rho} (\hat D^\rho \phi_2\phi_2^*-\hat D^\rho \phi_1 \phi_1^*-h.c.)}{2 q |\phi|^2}\label{dualA}\\
&+&\frac{(\hat D^\mu \phi_2\phi_2^*-\hat D^\mu \phi_1\phi_1^*-h.c.)^2}{4 |\phi|^2}.\nn
\end{eqnarray}

The two gauge $U(1)$ transformations
can be used to fix the components $\rho=0,1$ of $\hat D^\rho \phi_1 \phi_1^*$ 
to be real leading to the Lagrangian
{\color{red} }
\begin{eqnarray}
L_1&=&(\hat D_\mu \phi_1)^\dagger \hat D^\mu \phi_1+(\hat D_\mu \phi_2)^\dagger \hat D^\mu \phi_2+\frac{\partial_\mu \lambda \partial^\mu \lambda}{4 q^2|\phi|^2}-\frac{i \partial^\mu\lambda  \epsilon_{\mu\rho} (\hat D^\rho \phi_2\phi_2^*-h.c.)}{2 q |\phi|^2}\\
&+&\frac{(\hat D^\mu \phi_2\phi_2^*-h.c.)^2}{4 |\phi|^2}.\nn
\end{eqnarray}

Let us write explicitly $\phi_1=r_1e^{i\theta_1}$ to obtain that 
\begin{eqnarray}
(\hat D_\mu \phi_1)^\dagger \hat D^\mu \phi_1=(\partial_{\mu}r_1)^2+r_1^2(\partial_\mu \theta_1+\hat A_{\mu})^2.\nn
\end{eqnarray}

Results of (\ref{dualA}) can be used to analyze two other simpler cases:

\begin{enumerate}
\item
 When we keep only one scalar field $\phi_1$ and no gauge field such that $\hat D_{\mu}=\partial_\mu$, the gauge symmetry can be used to eliminate the phase
of $\phi_1$ leaving the Lagrangian
\begin{eqnarray}
L_1=\partial_\mu \phi_1 \partial^\mu \phi_1+\frac{(\partial_0 \lambda)^2}{4 q^2|\phi|^2},
\end{eqnarray}
with $\phi_1=\phi_1^*$ and $\lambda$ the dual component.

\item When we keep the two scalar fields $\phi_1$ and no gauge field, 
the gauged symmetry can be used to eliminate one phase.
Choosing the one of $\phi_1$ one obtains the Lagrangian
\begin{eqnarray}
L_1&=&\partial_\mu \phi_1 \partial^\mu \phi_1+\partial_\mu \phi_2^* \partial^\mu \phi_2+\frac{(\partial_0 \lambda)^2}{4 q^2|\phi|^2}
-\frac{i \partial^\mu\lambda  \epsilon_{\mu\rho} (\hat \partial^\rho \phi_2\phi_2^*-h.c.)}{2 q |\phi|^2} \\
&+&\frac{(\partial^\mu \phi_2\phi_2^*-h.c.)^2}{4 |\phi|^2}.\nn
\end{eqnarray}
with $\phi_1=\phi_1^*$ and $\lambda$ the dual component.

In this Appendix we have presented the main elements of a non-Abelian $SU(2)$ duality which renders
the scalar  part of a GLSM dualization along $\sigma_3$ direction. Further analysis of this connection
will be presented elsewhere.

\end{enumerate}

\bibliographystyle{utphys}
\bibliography{biblioNatd}

\end{document}